\documentclass[english,american]{elsarticle}
\usepackage[LGR,T1]{fontenc}
\usepackage[latin9]{inputenc}
\usepackage{float}
\usepackage{amsmath}
\usepackage{graphicx}
\usepackage{setspace}
\usepackage{esint}
\usepackage{subscript}
\usepackage{color}

\makeatletter

\providecommand{\tabularnewline}{\\}

\usepackage{fullpage}

\usepackage{babel}
\newcommand{\ignore}[1]{}

\makeatother

\usepackage{babel}
\begin{document}

\begin{frontmatter}{}

\title{Level-set method for accurate modeling of two-phase immiscible flow with moving contact lines}

\author{Moataz O. Abu-Al-Saud\textsuperscript{a}, Cyprien Soulaine\textsuperscript{a}, Amir Riaz\textsuperscript{b}, Hamdi A. Tchelepi\textsuperscript{a}}

\address{\textcolor{black}{\textsuperscript{a}Department of Energy Resources Engineering, Stanford
University, Stanford, CA 94305, USA}}
\address{\textcolor{black}{\textsuperscript{b}Department of Mechanical Engineering, University of Maryland , College Park, MD 20742, USA}}

\begin{abstract}
We developed a sharp interface level-set approach for two-phase immiscible
flow with moving contact lines. The Cox-Voinov model is used to describe
the moving contact line. A piecewise linear interface method is used
to construct the signed distance function and to implement the contact
angle boundary condition. Spurious currents are studied in the case
of static and moving fluid interfaces, which show convergence behavior.
Pressure and the surface tension force are balanced up to machine
precision for static parabolic interfaces, while the velocity error
decreases steadily with grid refinement when the interface is advected
in a uniform flow field. The moving contact line problem is studied
and validated through comparison with theory and experiments for an
advancing interface and capillary rise in a tube.
\end{abstract}

\end{frontmatter}{}

\section{Introduction}\label{2.1}

Two-phase flows with moving contact lines (MCL) occur in a wide range
of physical process. Examples include two-phase flow in porous media
(enhanced oil recovery \cite{lake1989enhanced}, CO\textsubscript{2}
sequestration \cite{cinar2014carbon}, and fuel cells \cite{He2000}),
coating flows \cite{kistler1997liquid}, and ink-jet printing \cite{2015Kumar}.
There are several challenges in modeling multiphase flows with moving
contact lines from both the physical and numerical standpoints. The
physics of such flows involves effects that originate at the molecular
scale \cite{2013Snoeijer}. The conventional continuum hydrodynamic
approach based on the no-slip assumption is inappropriate because
it implies the presence of an unbounded stress singularity \cite{HUH1971}.
In terms of numerical simulation, the nano-meter length-scale associated
with the MCL is computationally impractical to resolve for problems
that have micro-meter or milli-meter characteristic length-scale. In
addition, the numerical error in computing surface tension force can
severely interfere with the simulation results. The so-called spurious
currents associated with errors in curvature computations remain a
major obstacle for low capillary number problems, such as two-phase
flow in porous media. The capillary number in two-phase flow in porous
media, defined as ${Ca=\frac{\mu u}{\gamma}}$ where ${\mu}$, ${\gamma}$,
and ${u}$ are viscosity, surface tension, and velocity respectively,
can range from ${10^{-3}}$ to ${10^{-9}}$ \cite{lake1989enhanced}.

Moving contact line (MCL) models can be implemented using various
interface advection schemes, such as front tracking \cite{Tryg1},
diffuse interface \cite{Jacqmin1}, volume-of-fluid (VOF) \cite{Hirt1},
and level-set method \cite{Sussman1} (see \cite{Sui1} for recent
review). A straight-forward approach is to implement a contact angle
boundary condition with no-slip velocity at the wall. The wall is
typically taken to coincide with the cell boundary while the velocity
component parallel to the wall is defined at the cell-center. Grid
convergence cannot be achieved with this approach because of the absence
of a slip model. Thus, while no-slip velocity condition may apparently
work for coarser grids, it does not converge and becomes numerically
intractable with grid refinement. As a remedy, molecular interactions
in the vicinity of the contact-line must be considered to account
for the correct physics of moving contact lines. Several phenomenological
approaches are available that employ physically relevant microscopic
parameters to account for the underlying molecular effects. These
include interface diffusion model \cite{rowlinson2002molecular},
the precursor film model based on Van der Waals forces \cite{DeGennes1985},
and slip-based models \cite{1979Dussan}. The parameters associated
with such models are typically on the order of nano-meters, and that
places severe constraints on our ability to resolve the physics for
problems of practical interest.

Theoretical hydrodynamic models for the dynamic contact line have
been developed using the technique of matched asymptotic expansions
to relate the observed macroscopic contact angle to the nano-scale
region. The results based on these models are found to be in good
agreement with experimental results for small values of the capillary
and Reynolds numbers \cite{Hoffman1975,STROM1990}. In this context,
the Cox-Voinov \cite{Cox1,Voinov1976} relation is a slip-based model
according to which the macroscopic contact angle deviates from the
microscopic angle mainly due to the phenomenon of viscous bending
that occurs at an intermediate length-scale, between the molecular
and the capillary length scales. The latter can be regarded as the
macroscopic scale related to meniscus curvature. The Cox-Voinov model
has been implemented in VOF and level-set frameworks \cite{Sui2013,Dupont2010,Afkhami2009,afkhami2017transition}
using the Continuous Surface Force (CSF) method \cite{Brackbill1992}.

% \cite{Brackbill1992}\textcolor{red}{and
% the delta-function models for the implementation of surface tension
% forcing at the interface. Both the CSF and the delta function methods
% use the smeared interface approach, which can produce substantial
% unphysical vorticity, or spurious currents, at the interface \cite{Abadie2015}.
% Moreover, the standard delta-function implementation is also inconsistent
% at the interface \cite{Engquist2005}.} 

Interface advection with the VOF method conserves mass. However,
VOF curvature estimation can be tedious because VOF is a discontinuous
function. Iterative smoothing as well as filtering techniques have
been applied to improve the curvature computation in VOF \cite{Lafaurie1994,Raeini2012}.
However, both smoothing and filtering have not been able to mitigate
the problem of spurious currents with grid refinement for problems
involving contact angles \cite{Dupont2010,Raeini2012}. In addition,
the filtering technique used in \cite{Raeini2012} may potentially
suppress physical capillary waves on the interface that have been
shown to play a significant role in how interfaces move in confined
domains \cite{Roman2015,Moebius2014}. Curvature computation of
VOF has been further improved using height functions. A perfect numerical
balance between surface tension force and pressure gradient has been
demonstrated for VOF using height functions and piecewise linear reconstruction
(PLIC) for the static droplet case \cite{Francois1}; however, spurious
currents still creep in and magnify for moving droplets \cite{Popinet2009,Abadie2015}.
The height function method has been extended to include contact lines
\cite{Afkhami2007}. While it is shown to be grid converging, spurious
currents could not be eliminated for static droplets on solid surfaces.
In addition, the interface reconstruction from VOF information is
computationally expensive in 3D \cite{Wu2013}. The improvement of
VOF with regards to curvature is an active area of research \cite{owkes2015}.

In this study we use a level-set approach for tracking the interface,
separating the two immiscible fluids. When the level-set function
is chosen to be a signed distance function, it leads to more accurate
and convenient curvature estimation compared with VOF, especially when the interface is advected \cite{Abadie2015}. The work in
\cite{spelt2005level} used level-sets to solve MCL problems. However,
the surface tension force is implemented using the CSF-formulation
which suffers from spurious currents that do not converge with grid
refinement. The new approach developed in this work is based on the
sharp treatment of the contact line model using the ghost-fluid method
\cite{Liu2000}. We demonstrate that with this new approach, spurious
currents are on the order of machine precision for static interfaces
involving contact angle boundary condition. Moreover, the spurious
currents are mitigated for both static and moving interface in a constant
flow field. This feature is very important in low capillary number
flows to avoid spurious currents that can be of the same order of
magnitude, or even greater than the physical flow field.

% \textcolor{red}{One drawback
% in using level-set is its susceptibility to lose mass due to numerical
% diffusion error during advection, and interface perturbation due to
% reinitialization. However, \cite{Qin2015} recently showed that most
% of the mass loss occurs in the conventional Hamilton-Jacobi (PDE-based)
% reinitialization of the level-set. Moreover, the reinitialization
% based on reconstruction of the implicit interface proposed in \cite{Qin2015}
% with high-order advection schemes \cite{Jiang2000,JIANG1996} results in significant
% improvement in mass conservation comparable to coupled Level-Set Volume
% of Fluid (CLSVOF) in \cite{Wang2012}.} % occurring in small-scale flows.

In the present work, the level-set method with projection piecewise
linear interface reconstruction and MCL model has been implemented.
The level-set and governing equations are introduced in section \ref{2.2}.
The MCL model and level-set reinitialization with contact angle boundary
condition are explained in section \ref{2.3}. Section \ref{2.4} includes verification
results as well as validation of the MCL model based on comparison
with theory and experiments such as flow in capillary tube, and capillary
rise.

\section{Level-set and two-phase flow equations}\label{2.2}

The level-set function ${\phi(\mathbf{x},t)}$ is used to capture
the fluid interface, which is an implicit function equal to zero at
the interface, positive and negative for each fluid respectively.
It is advected by:

\begin{equation}
\frac{\partial\phi}{\partial t}+\mathbf{u}\cdot\nabla\phi=0,\label{eq:LSadvection}
\end{equation}
where ${\mathbf{u}}$ is the external divergence-free velocity corresponding
to incompressible Navier-Stokes equations. The level-set field is
transferred to a signed distance function, while preserving the interface
location through projection piecewise linear reconstruction method
described in section \ref{2.3.2} with contact angle boundary condition implementation.

Assuming both fluids are immiscible and incompressible under isothermal
conditions, the governing equations are:

\begin{equation}
\nabla\cdot\mathbf{u}=0,\label{eq:continuity}
\end{equation}
and

\begin{equation}
\rho\bigg[\frac{\partial\mathbf{u}}{\partial t}+(\mathbf{u}\cdot\nabla)\mathbf{u}\bigg]=-\nabla p+\nabla\cdot\bigg(\mu\big(\nabla\mathbf{u}+\nabla\mathbf{u}^{T}\big)\bigg)+\rho\mathbf{g},\label{eq:momentum}
\end{equation}
where ${\rho}$ is the density, ${\mathbf{u}}$ is the flow velocity
vector, ${t}$ is time, ${p}$ is pressure, ${\mu}$ is viscosity,
and ${\mathbf{g}}$ is gravity vector in the y-direction. Because there
are two fluid phases, fluid interface boundary conditions have to
be satisfied. Based on level-set values, fluid densities, viscosities,
normal and curvature interfaces are computed. The densities and viscosities
are computed using a transition function:

\begin{equation}
\rho(\phi)=\rho_{1}+(\rho_{2}-\rho_{1})I(\phi),\label{eq:rho}
\end{equation}
and

\begin{equation}
\mu(\phi)=\mu_{1}+(\mu_{2}-\mu_{1})I(\phi),\label{eq:viscosity}
\end{equation}
where the subscripts 1 and 2 indicate the corresponding fluid phase.
${I(\phi)}$ is a smooth transition function based on the level-set
values defined as:

\begin{equation}
I(\phi)=\frac{1}{2}[1-\textrm{erf}(\phi/\epsilon)],\label{eq:smoothing}
\end{equation}
where ${\mathbf{\textrm{erf}}}$ is the error-function that smoothly
goes from -1 to 1 when it changes signs from negative to positive,
and ${\epsilon}$ is the width of the interface transition, which
is taken to be 1.5 times the grid size. The sharp transition in densities
and viscosities between the two-phases has been tested, but smoothing
the densities and viscosities using the error-function has been found
to be more stable. When the viscosity is smeared, the normal-stress
jump condition at the fluid interface becomes \cite{Kang2000}:

\begin{equation}
[p]_{I}=\gamma\kappa,\label{eq:Pjump}
\end{equation}
where ${\kappa}$ and ${\gamma}$ are the interface curvature and
surface tension. The brackets ${[.]_{I}}$ denote the sharp jump across
the interface. The curvature of the interface is computed from the
level-set:

\begin{equation}
\kappa=\nabla\cdot\frac{\nabla\phi}{|\nabla\phi|}.\label{eq:curvature}
\end{equation}
The numerical framework used in solving the above equations is structured
Marker and Cell (MAC) method \cite{harlow1965numerical}, where the velocity vector field is defined
at the grid interfaces, and the scalar fields are defined at the grid
center. The continuity and momentum equations are solved using the projection
method \cite{Chorin1967}. The advection of the level-set function, Eq.\
\ref{eq:LSadvection}, is discretized using a fifth-order WENO scheme \cite{JIANG1996}.
The pressure jump across the fluid interface, Eq.\ \ref{eq:Pjump}, is implemented using the ghost-fluid method \cite{Liu2000}.
The curvature is evaluated using central-differences. For grid cells
near the wall, the curvature computation is illustrated in section
\ref{2.3}.

\section{Moving contact line and level-set reinitialization}\label{2.3}

\subsection{Moving contact line model}\label{2.3.1}

To understand and predict multiphase flow in confined spaces, such
as porous media and micro-channels, moving contact lines (MCL) need
to be modeled accurately. This is because MCLs determine the fluid
interface shape, as well as the distribution of each fluid phase with
respect to the solid surface. The physics of problems involving MCLs
is multi-scale. For instance, the characteristic length-scale of a
channel or pore-width can be 100 micron whereas MCL is a molecular
process on the order of a nano-meter. As a result, it is impractical
to resolve the nano-meter scale to model a flow with MCL in a domain
size of milli-meter or micro-meter. Previous results \cite{Legendre2014}
show that numerical results can vary significantly if the equilibrium
contact angle based on Young's equation is implemented as a boundary
condition without resolving the nano-meter scale.

\begin{figure}[h]
\centering{} \includegraphics[width=0.9\textwidth]{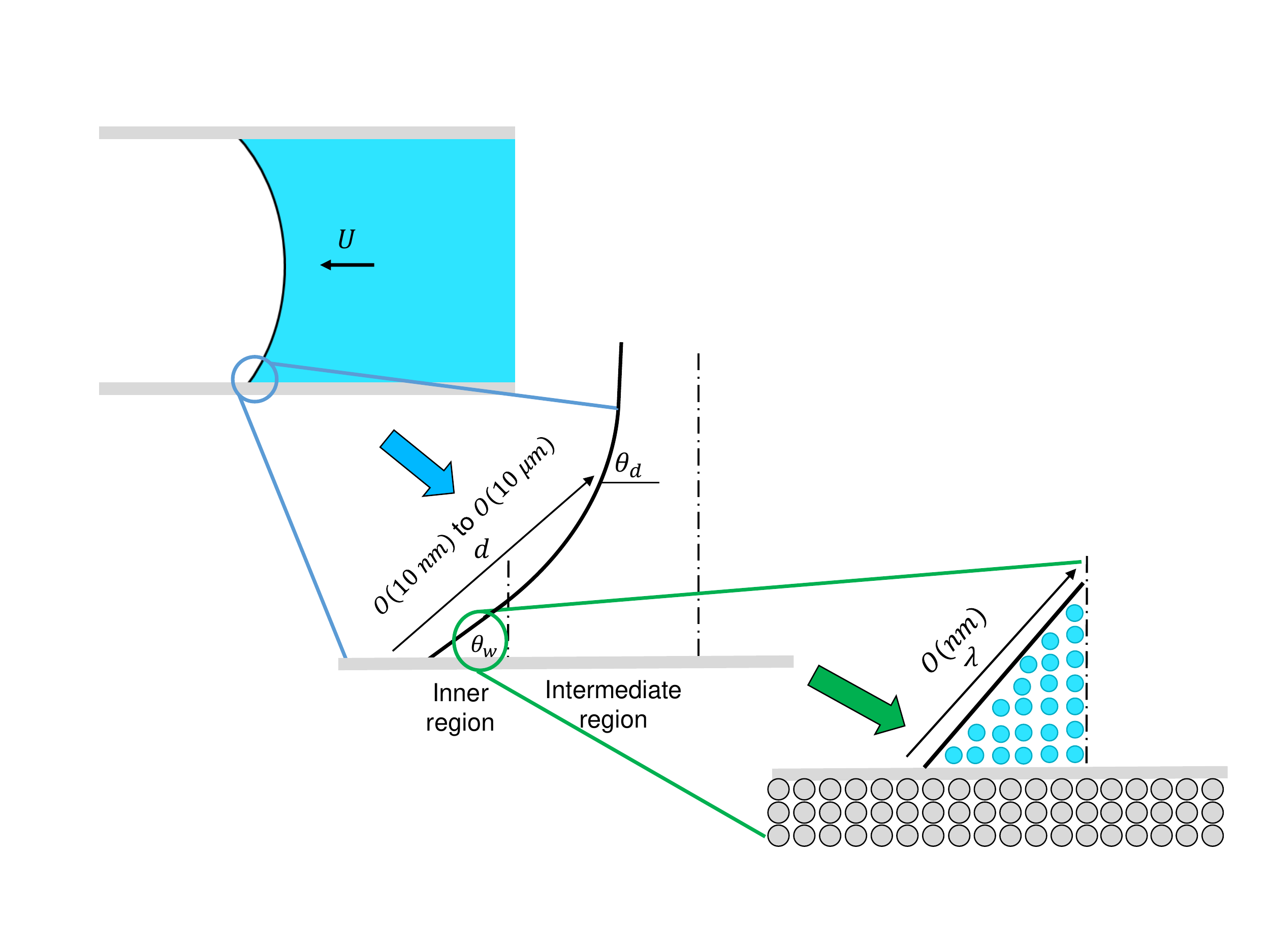}
\caption{Length scales involved in flows with moving contact line\label{fig:MCL_scales}}
\end{figure}
Hydrodynamic theories have been developed to understand the MCL problem
via matched asymptotic techniques. Figure \ref{fig:MCL_scales} illustrates the length
scales that are involved in the MCL problem. In the method of matched
asymptotic expansions, the solution within the inner region near the
contact line is matched to the outer region through an intermediate
region which is the region that is geometry-independent and where
viscous and capillary forces are of comparable magnitude. With the
assumption of low capillary numbers (${Ca\le0.1}$) and smooth flat solid surface, Cox \cite{Cox1}
developed a theory that relates the contact angle  at the intermediate
region to the inner-scale contact angle:
% \ignore{, negligible viscosity of fluid 2 in figure}
\begin{equation}
g(\theta_{d},\mu_{2}/\mu_{1})=g(\theta_{w},\mu_{2}/\mu_{1})+\text{Ca}_{d}\ln\bigg(\frac{d}{\lambda}\bigg)+\text{Ca}_{d}Q_{int},\label{eq:Cox}
\end{equation}
where the function ${g}$ is expressed as:

\begin{equation}
g(\theta,\mu_{2}/\mu_{1})=\int_{0}^{\beta}\bigg(\frac{d\beta}{f(\beta,\mu_{2}/\mu_{1})}\bigg)%\frac{1}{2}\int_{0}^{\theta}\bigg(\frac{\theta-\sin\theta\cos\theta}{\sin\theta}\bigg)d\theta,{equation},
,\label{eq:g_func}
\end{equation}
and

\begin{equation*}
f(\beta,\mu_{2}/\mu_{1})=\frac{2\sin\beta\bigg[\big(\mu_{2}/\mu_{1}\big)^{2}\big(\beta^{2}-\sin^{2}\beta\big)+2\big(\mu_{2}/\mu_{1}\big)\big[\beta(\pi-\beta)+\sin^{2}\beta\big]+\big[(\pi-\beta)^{2}-\sin^{2}\beta\big]\bigg]}{\big(\mu_{2}/\mu_{1}\big)\big(\beta^{2}-\sin^{2}\beta\big)\big[(\pi-\beta)+\sin\beta\cos\beta\big]+\big[(\pi-\beta)^{2}-\sin^{2}\beta\big](\beta-\sin\beta\cos\beta\big)}%\frac{1}{2}\int_{0}^{\theta}\bigg(\frac{\theta-\sin\theta\cos\theta}{\sin\theta}\bigg)d\theta{equation}
.\label{eq:f_func}
\end{equation*}
In equation \ref{eq:Cox}, ${\mu_{1}}$ is the viscosity of the displacing fluid,
${\mu_{2}}$ is the viscosity of the displaced fluid, ${\theta_{d}}$
is the contact angle at the intermediate scale, ${\theta_{w}}$ is
the contact angle within the inner region, ${\text{Ca}_{d}}$ is the
capillary number corresponding to the intermediate scale, ${d}$ is
the radial distance from the contact line triple point to the intermediate
contact angle ${\theta_{d}}$, ${\lambda}$ is the cut-off length
scale determining the limit of continuum hydrodynamics, and ${Q_{int}}$
is a higher-order approximation term on the order of $O(Ca)$
\cite{Cox1}. ${\lambda}$ is a physical parameter corresponding
to a molecular process, which both Cox \cite{Cox1} and Voinov \cite{Voinov1976}
assumed to be the slip length, and it is at the order of few times
the size of a fluid molecule. ${\theta_{w}}$ is still difficult to
experimentally measure, but is usually assumed to be ${\theta_{eq}}$
for viscous flows \cite{Bonn2009}. The assumption of $\theta_{w}=\theta_{eq}$
has been supported by various experiments \cite{Hoffman1975,STROM1990}.
${Q_{int}}$ is an asymptotic ${O(Ca)}$ term that depends on the
slip-length ${\lambda}$, viscosities ${\mu_{2}/\mu_{1}}$, and ${\theta_{w}}$,
which is important for matching asymptotic theory with numerical results
as shown in \cite{Sui2013}. Hocking and Rivers \cite{Hocking1982}
computed the second order leading term (${Q_{i}}$ which is ${Q_{int}-1}$)
based on ${\theta_{w}}$ and the assumption of slip mechanism and
negligible ${\mu_{2}}$. When the viscosity of the displacing fluid
is much larger than that of the displaced fluid, i.e. negligible ${\mu_{2}/\mu_{1}}$,
equation \ref{eq:g_func} simplifies to:

\begin{equation}
g(\theta)=\frac{1}{2}\int_{0}^{\beta}\bigg(\frac{\beta-\sin\beta\cos\beta}{\sin\beta}\bigg)d\beta.\label{eq:g_func2}
\end{equation}
Moreover, when ${\theta_{d}\le3\pi/4}$ equation \ref{eq:Cox} becomes \cite{Voinov1976}:

\begin{equation}
\theta_{d}^{3}=\theta_{eq}^{3}+Ca_{d}\bigg[9\bigg(\ln\frac{d}{\lambda}\bigg)+Q_{int}\bigg].\label{eq:Voinov}
\end{equation}
The above equation describes the relation between the intermediate
contact angle, ${\theta_{d}}$, and ${\theta_{eq}}$ for ${Ca_{d}\le0.1}$,
smooth surface, small ${\mu_{2}/\mu_{1}}$, and ${\theta_{d}\le3\pi/4}$.

\begin{figure}[H]
\centering{} \includegraphics[width=0.9\textwidth]{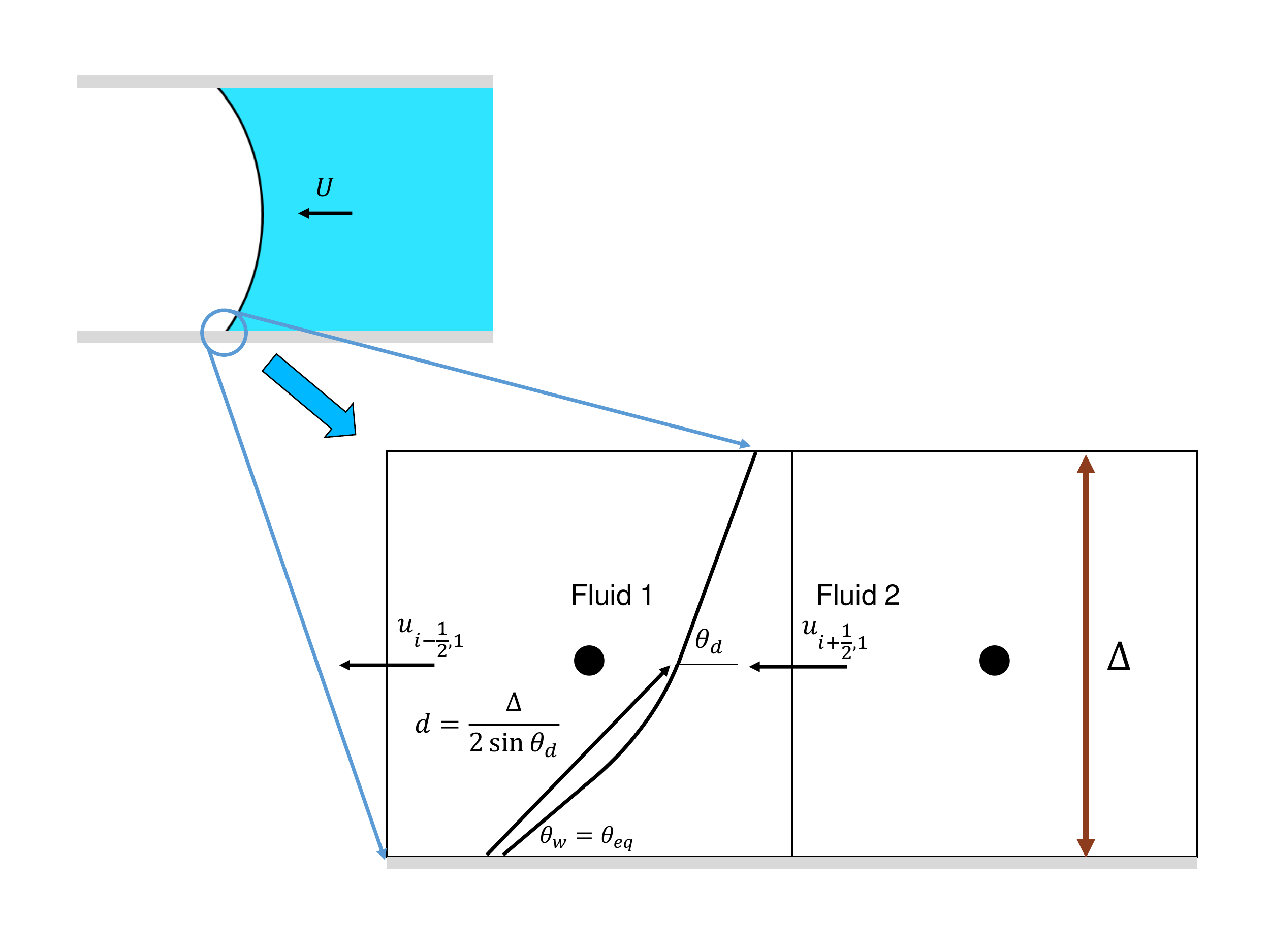}
\caption{The implementation of dynamic contact angle in grid points near the
contact line\label{fig:MCL_grid}}
\end{figure}
The numerical implementation of equation \ref{eq:Voinov} is illustrated in Figure \ref{fig:MCL_grid}. The local capillary number, $Ca_{d}=\mu U_{d}/\gamma,$ is computed
by averaging the parallel wall velocities in the cell containing the
contact line, that is $U_{d}=(u_{i-1/2,1}+u_{i+1/2,1})/2$. We have
observed that using this form of the averaged velocity to determine
the local capillary number produces less spurious currents compared
with a linear interpolation of velocity at the interface.  The radial
distance, $d$, is calculated as shown in Figure \ref{fig:MCL_grid} which depends on
the grid size and corresponds to the intermediate length scale. After
$\theta_{d}$ is determined, the normal vector in the cells neighboring
the contact line is computed via:

\begin{eqnarray}
\vec{n} & = & \vec{n}_{||}\sin\theta_{d}+\vec{n}_{\perp}\cos\theta_{d},\label{eq:normal}
\end{eqnarray}
where $\vec{n}_{||}$ and $\vec{n}_{\perp}$ are parallel and normal
solid surface vectors. Once the normal vector is determined, the curvature
is computed 

\begin{equation}
\kappa=\nabla\cdot\vec{n}=\frac{\partial n_{x}}{\partial x}+\frac{\partial n_{y}}{\partial y}.\label{eq:curv_comp}
\end{equation}
Equation \ref{eq:curv_comp} is discretized using central differences as mentioned
in section \ref{2.2}. However, for grid cells near the wall, the derivative
of the normal in the direction perpendicular to the wall is discretized
using a one-sided stencil. The one-sided finite-difference provided
more accurate results in terms of both spurious currents and curvature
when it was compared with central finite difference which requires
defining ghost cell points below the wall. The next section illustrates
the reinitialization of level-set field and how contact angle boundary
condition is taken into consideration.

\subsection{Level-set reinitialization with contact angles}\label{2.3.2}

\begin{figure}[H]
\centering \includegraphics[width=0.8\textwidth]{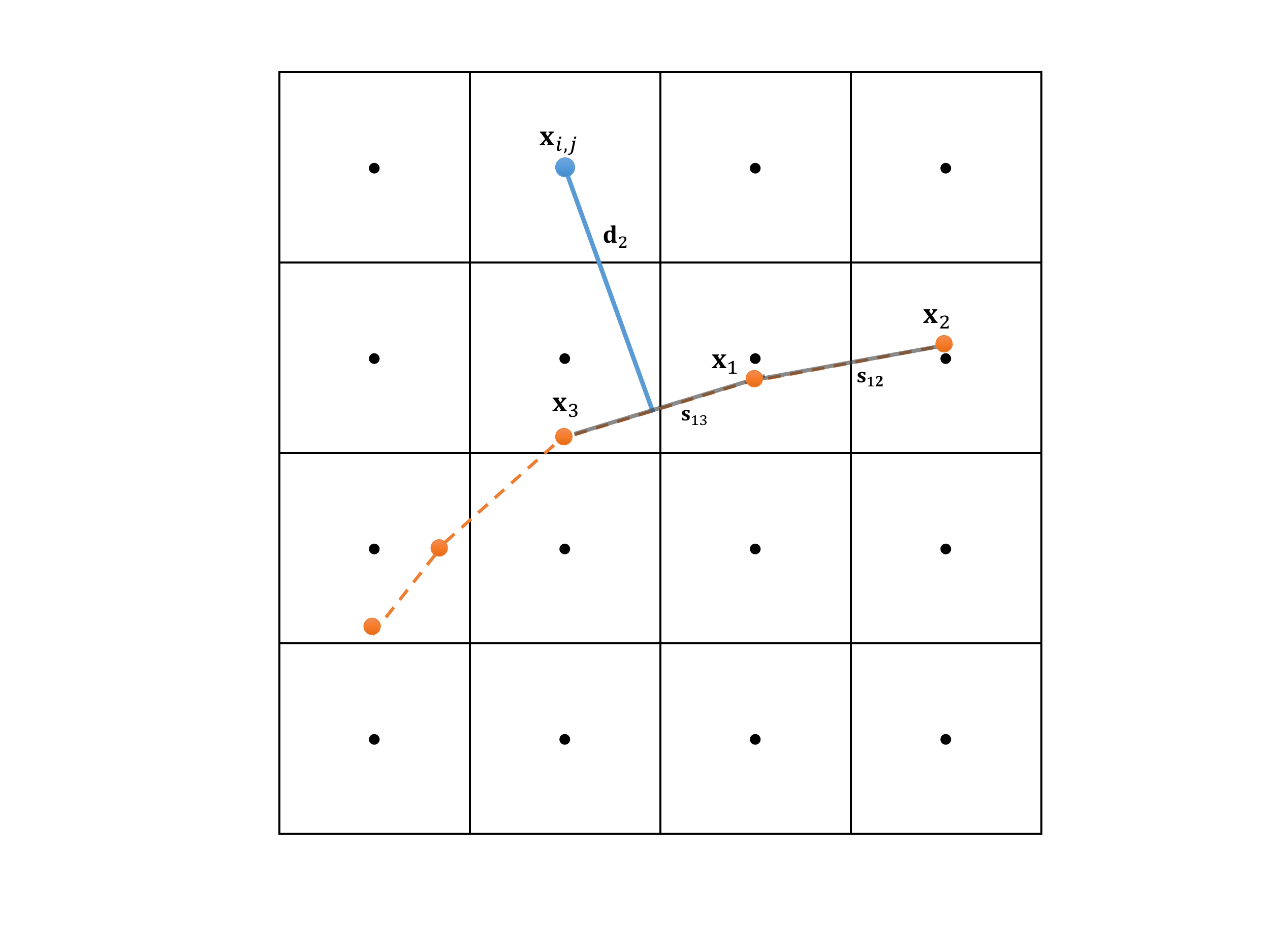}
\caption{Piecewise linear projection to compute the distance from a grid point
to a reconstructed piecewise linear function}
\label{fig:project_LS} 
\end{figure}

Computing the signed distance function based on point-wise, piecewise
linear, and smooth reconstruction of the interface from the level-set
function is elaborated in details in \cite{Qin2015}. In our numerical
method, piecewise linear interface reconstruction (called P\textsubscript{2}
in \cite{Qin2015}) is used to compute signed distance function.
First, the discrete interface points corresponding to ${f(\mathbf{x})=0}$
is found from level-set. The discrete interface points are determined
when values of the level-set defined on grid points change sign in
x and y directions. Then, three discrete points that have the shortest
distances to the grid point are determined (see Figure \ref{fig:project_LS}):

\begin{equation}
|\mathbf{x}_{i,j}-\mathbf{x}_{1}|<|\mathbf{x}_{i,j}-\mathbf{x}_{2}|<|\mathbf{x}_{i,j}-\mathbf{x}_{3}|.\label{eq:distance}
\end{equation}

Afterward, the minimal distance ${\mathbf{d}_{2}=d_{i,j}}$ to the
piecewise linear reconstructed interface from the three points ${\mathbf{x}_{1},\mathbf{x}_{2}}$,
and ${\mathbf{x}_{3}}$ is computed through normal projection of $\mathbf{x}_{i,j}$.
The minimal distance ${\mathbf{d}_{2}}$ normally intersects one of
the segments ${s_{13}}$ or ${s_{12}}$. The normal distance from
a point ${x_{i,j}}$ with coordinates ${(x_{i},y_{i})}$ to a line
containing two points ${\mathbf{x}_{1}}$ and ${\mathbf{x}_{3}}$,
with coordinates ${(x_{1},y_{1})}$ and ${(x_{3},y_{3})}$ respectively,
is:

\begin{equation}
\mathbf{d}_{2}=\frac{|(x_{i}-x_{3})(y_{1}-y_{i})-(x_{i}-x_{1})(y_{3}-y_{1})|}{\sqrt{(y_{3}-y_{1})^{2}+(x_{3}-x_{1})^{2}}}.\label{eq:distance2}
\end{equation}
The above distance formula is applied for each segment ${s_{12}}$
and ${s_{13}}$. In Figure \ref{fig:project_LS}, ${\mathbf{d}_{2}}$ normally intersects
${s_{13}}$ such that ${\mathbf{d}_{2}\cdot\mathbf{t}_{13}=0}$, where
${t_{13}}$ is the tangent vector to the segment ${s_{13}}$.

\begin{figure}[H]
\centering \includegraphics[width=0.6\textwidth]{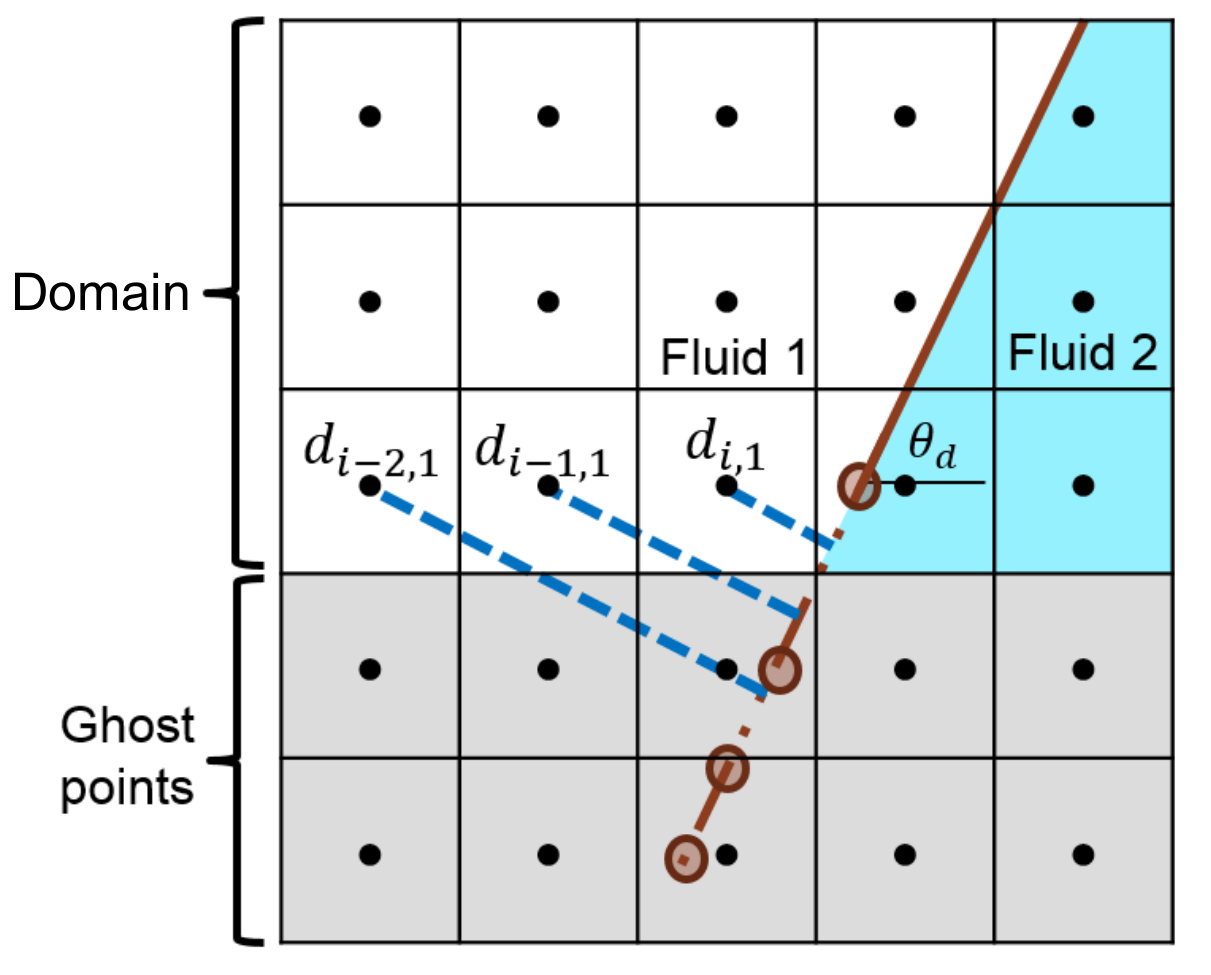}
\caption{Piecewise linear projection to compute the distance from a grid point
to a reconstructed piecewise linear function with contact angle boundary
condition}
\label{fig:project_contactangle} 
\end{figure}

When the interface intersects the solid surface, a contact angle boundary
condition needs to be imposed. Figure \ref{fig:project_contactangle} shows a schematic of how the
distance function is constructed and computed when the interface intersects
the solid boundary at a specified contact angle. The interface is
extended linearly with a slope determined by the contact angle, so
that reinitialization and distance computation is done correctly for
grid points located near the contact line as illustrated in \cite{spelt2005level}.
The contact angle with the level-set method was implemented in the
framework of PDE-based reinitialization \cite{spelt2005level,DellaRocca2014}.
In our approach, the extension of the interface is accomplished by
linearly constructing the interface from the intersection points in
the ghost domain. For instance, the values of $d_{i-2,1}$, $d_{i-1,1}$,
and $d_{i,1}$ represent the respective distances from the linear
extension of the contact line beyond the wall. In the algorithm,
the level-set function is not defined in the ghost points as there
is no need to do so because the curvature is computed by one-sided
differences as explained in section \ref{2.3.1}. The location of the intersection
points along the x-direction (brown circles in Figure \ref{fig:project_contactangle}) are determined
via:

\begin{equation}
x_{lin,n}=x_{cl}-\left(\frac{\Delta}{\tan\theta}\right)k,\label{eq:xpoint}
\end{equation}
and

\begin{equation}
y_{lin,n}=y_{i}-\left(\Delta\right)k,\label{eq:ypoint}
\end{equation}
where ${x_{cl}}$ is the contact line defined at the grid-scale, and
${k}$ is an index between 1 and the number of intersection points
in the x-direction. In the y-direction, the location of the intersection
points (blue circle in Figure \ref{fig:project_contactangle}) are:

\begin{equation}
x_{lin,n}=x_{i+1-k}%x_{cl}-(n)\frac{\Delta}{\tan\theta}
,\label{eq:xpoint2}
\end{equation}
and

\begin{equation}
y_{lin,n}=(x_{cl}-x_{i+1-k})\tan{\theta},\label{eq:ypoint2}%y_{i}-(n)\Delta.{equation}
.
\end{equation}
Because curvature calculation involves a 5-point stencil, the minimum
number of intersection points in each direction is three (which is
the number that has been used in section \ref{2.4}) to have correct reinitialization
for cells near the contact line. Once the interface is extended along
the contact angle and the intersection points are determined, the
distance values at grid points inside the domain are computed via
piecewise linear interface reconstruction as in equation \ref{eq:distance2}.

\section{Results}\label{2.4}

This section describes the results obtained using the new approach
presented above. We start by first considering the performance of
the method for problems that do not involve contact angles, for which
the accuracy of piecewise linear interface projection is investigated
with the help of static and oscillating droplet problems.

\subsection{Static droplet}\label{2.4.1}

\begin{figure}
\centering{} \includegraphics[width=0.65\textwidth]{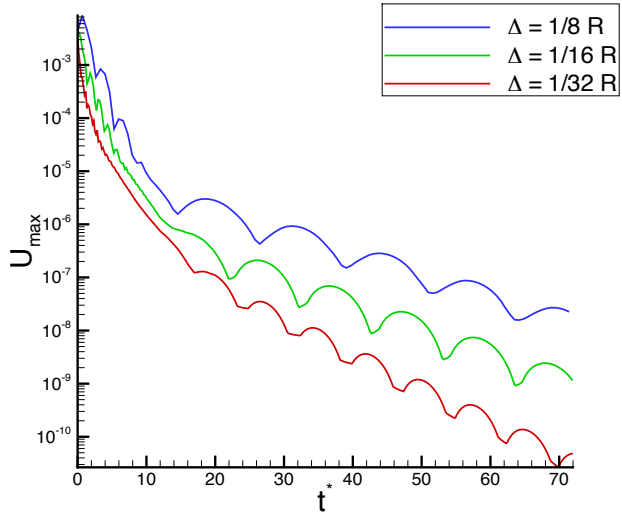}
\caption{Velocity convergence to zero for static droplet at different grid
sizes}\label{fig:umx_static_drop}
\end{figure}

In the static droplet problem, we test whether our new method allows
the spurious currents to be on the order of machine precision, indicating
an exact balance between the surface tension force and the pressure
jump. It is well known \cite{Abadie2015} that such an exact balance
cannot be achieved when the Hamilton-Jacobi reinitialization (PDE-based)
procedure is used together with the ghost-fluid method. The balance
between the pressure jump across the droplet and the surface tension
force has been achieved only when reinitialization step is not applied
\cite{Abadie2015}. When PDE-based reinitialization is used, the
interface is slightly perturbed, and the reinitialization process
has to be modified at grid points by manually propagating characteristics
based on the interface location \cite{Russo2000}.  On the other
hand, the method of reinitialization used in this work allows the
surface tension and the pressure jump to be balanced to machine precision.
Figure \ref{fig:umx_static_drop} shows that with our approach, the maximum velocity converges
to machine precision for different grid sizes. The results have been
obtained by solving for the velocity field as well as the pressure,
and reinitialization is carried out at every time-step. The dimensionless
Laplace number, defined as the ratio of the Reynolds and capillary
numbers, is suitable for characterizing the flow behavior:

\begin{equation}
La=\frac{\rho L\gamma}{\mu^{2}}=\frac{Re}{Ca},\label{eq:La}
\end{equation}
where ${L}$ is the characteristic length-scale, which is equal to
the interface radius. In this test, La = 12000, which is the same
number as the static droplet test case conducted in \cite{Popinet2009}.
The capillary waves triggered by initial local numerical curvature
errors can be observed as periodic oscillations. Also, the initial
numerical error decreases with grid refinement. This example shows
how the level-set method with piecewise linear interface reconstruction
balances the pressure jump and the surface tension for the static
droplet case.

\subsection{Oscillating droplet}\label{2.4.2}

\begin{figure}[H]
\centering{}\includegraphics[width=0.85\textwidth]{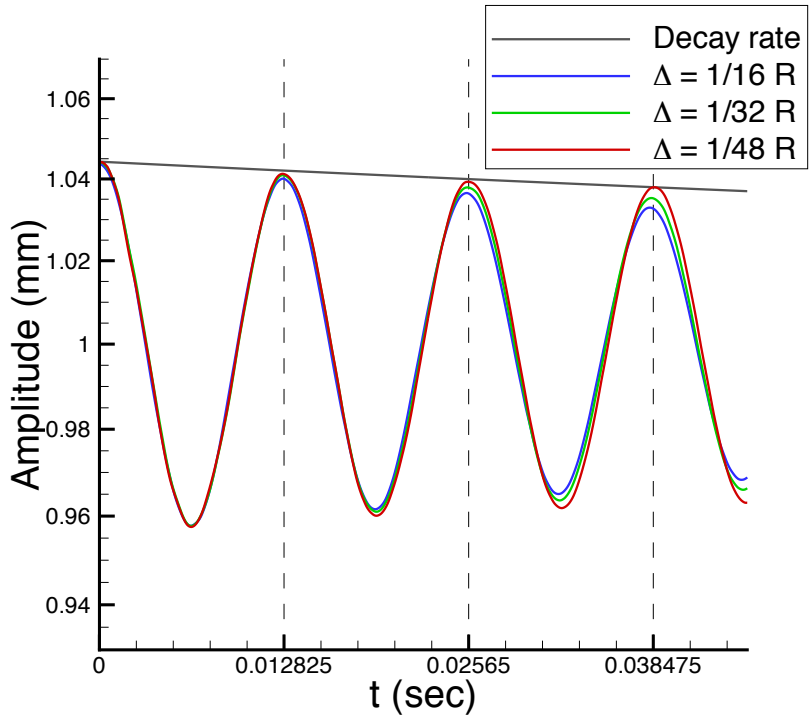}
\caption{Grid convergence of droplet amplitude to analytical solution for Re
= 200 and We = 1} \label{fig:amplitude}
\end{figure}

\begin{figure}[H]
\centering{}\includegraphics[width=0.85\textwidth]{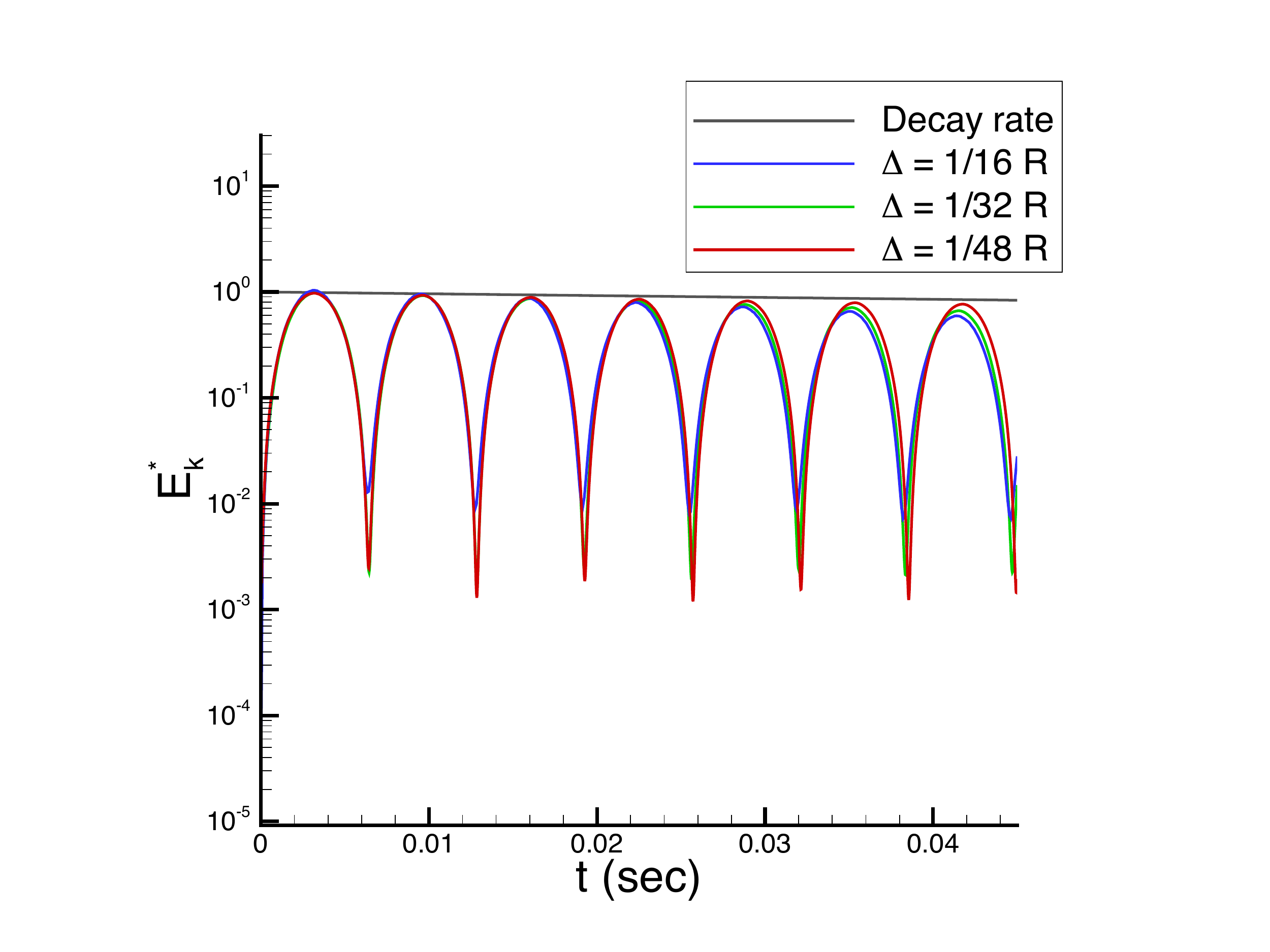}
\caption{Grid convergence of droplet energy to analytical solution for Re =
200 and We =1} \label{fig:energy}
\end{figure}

%Here, I will show the oscillating droplet example and compare between the conventional and P\textsubscript{2} projection
When the droplet is slightly perturbed, the physical imbalance between
the pressure jump and the surface tension gives rise to a restoring
flow field that tends to drive the droplet back towards its equilibrium
configuration. Lamb \cite{lamb1916hydrodynamics} derived an analytical
solution for the frequency and the amplitude of a perturbed viscous
droplet in vacuum. For a 2D droplet, the oscillation frequency of
the second mode $\omega_{2}$, is $\sqrt{\frac{6\gamma}{\rho R^{3}}}$
and the decay rate for both amplitude and kinetic energy is proportional
to $\exp\left(-\frac{4\mu}{\rho R^{2}}t\right)$. Figures \ref{fig:amplitude} and \ref{fig:energy}
show the droplet amplitude and kinetic energy when the amplitude is
perturbed by a factor of 1.04 relative to the equilibrium droplet.
The domain size is ${4R}\times$$4R$, and the problem is characterized
by Reynolds number $Re=\frac{\rho UL}{\mu}$ for the decay rate and
Weber number $We=\frac{\rho U^{2}L}{\gamma}$ for the oscillation
frequency. Table \ref{t1} lists the fluid parameters which correspond to
$We=1$ and $Re=200$. The Laplace number for this case is $La=40000$.
The viscosity and density ratio between the droplet and its surrounding
is 1/1000. Both the amplitude and the kinetic energy approach the
upper limit of the decay rate, and the time period of oscillation
matches the analytical solution. The solution becomes more accurate
as the grid is refined, and that shows that the numerical method converges
to the analytical solution.

% \textcolor{red}{We also compare with results obtained
% using second-order PDE-based reinitialization in Figures \ref{fig:amplitude2} and \ref{fig:energy2}.
% We observe that our reinitialization approach based on the piecewise
% linear interface projection produces more accurate results.  The
% proposed method agrees with the theoretical oscillation frequency,
% whereas the oscillation frequency in PDE-based reinitialization starts
% to deviate after a few oscillation cycles. Also, piecewise linear
% reconstruction conserves more energy as the Hamilton-Jacobi PDE-based
% reinitialization smoothens the level-set contours around the interface
% causing further damping of the energy. The piecewise linear reconstruction
% is also more efficient computationally. Table \ref{} shows the run times
% for both reinitialization methods for the oscillating droplet case
% where the entire domain is reinitialized.}

\begin{table}[H]
\begin{centering}
\begin{tabular}{|c|c|c|c|}
\hline 
R (mm) & $\rho$ (kg/m\textsuperscript{3}) & $\mu$ (cP) & $\gamma$(mN/m) \tabularnewline
\hline 
\hline
 1 & 1000 & 1 & 40 \tabularnewline
\hline
\end{tabular}
\par\end{centering}
\caption{Fluid properties corresponding to Re = 200 and We = 1}\label{t1}
\end{table}

%\begin{table}[h]
%\begin{centering}
%\begin{tabular}{|c|c|c|}
%\hline 
%${\Delta}$  & Piecewise projection & PDE-based\tabularnewline
%\hline 
%\hline 
%1/16  & 1.51 & 1.7\tabularnewline
%\hline 
%1/32 & 25.5 & 28.7\tabularnewline
%\hline 
%1/48 & 147.2 & 181.5\tabularnewline
%\hline \label{table_2.2}
%\end{tabular}
%\par\end{centering}
%
%\caption{Simulation time in minutes for piecewise projection and PDE-based
%reinitialization}\label{t2}
%\end{table}

%\begin{figure}[H]
%\centering{} \includegraphics[width=0.65\textwidth]{Amplitude_Re200_Freq.pdf}
%\caption{Droplet amplitude comparison between PDE-based reinitialization and
%piecewise projection for Re = 200 and We = 1 for $\Delta=$ 1/32 R} \label{fig:amplitude2}
%\end{figure}
%
%
%\begin{figure}[H]
%\centering{}\includegraphics[width=0.65\textwidth]{KineticEnergy_Re200.pdf}
%\caption{Droplet energy comparison between PDE-based reinitialization and piecewise
%projection for Re = 200 and We = 1 for $\Delta=$ 1/32 R}
%\label{fig:energy2} 
%\end{figure}

\subsection{Moving Wedge}\label{2.4.3}

\begin{figure}
\centering{}\centering \includegraphics[width=0.75\textwidth]{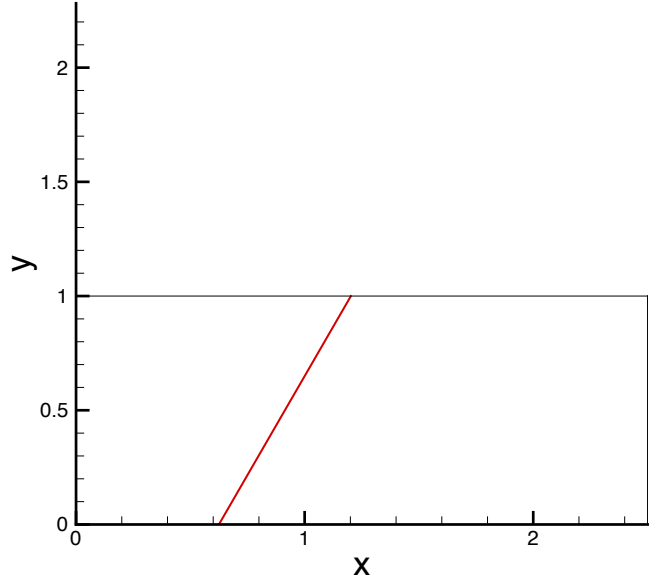}
\caption{Moving wedge: linear interface tilted at angle ${\theta=60}$}
\label{fig:wedge} 
\end{figure}
To test the piecewise linear projection method with the contact line
implementation, the translating wedge is tilted at an angle in uniform
flow from left to right as shown in Figure \ref{fig:wedge}. The wedge should maintain
its orientation with respect to the initially imposed angle, such
that the flow field is not disturbed. The numerical method, including
the coupling of Navier-Stokes, level-set advection, and piecewise
linear reconstruction is employed. The inflow capillary and Reynolds
number are ${Ca=0.01}$ and ${Re=1}$ respectively. Table \ref{t3} shows
velocity and curvature errors in ${L_{2}}$ and ${L_{\infty}}$ norms
at dimensionless time ${t^{*}=1}$, time normalized by ${L/U}$,
for different grid sizes where $L$ is the domain width. The grid
size ${\Delta}$ is the unit length of the domain divided by the number
of grid cells.  The velocity and curvature error norms have been
computed as follows:

\begin{equation}
||\mathbf{u}-\mathbf{\tilde{u}}||_{2}=\sqrt{\frac{1}{n}\sum_{i,j}(\mathbf{u}_{i,j}-\mathbf{\mathbf{\tilde{u}}}_{i,j})^{2}},\label{eq:u_l2}
\end{equation}

\begin{equation}
||\mathbf{u}-\mathbf{\tilde{u}}||_{\infty}=\max_{i,j}(|\mathbf{u}_{i,j}-\mathbf{\mathbf{\tilde{u}}}_{i,j}|),\label{eq:u_linf}
\end{equation}

\begin{equation}
||\kappa_{\Gamma}-\widetilde{\kappa}_{\Gamma}||_{2}=\sqrt{\frac{1}{n_{\Gamma}}\sum_{n_{\Gamma}}(\kappa_{\Gamma_{k}}-\widetilde{\kappa}_{\Gamma_{k}})^{2}},\label{eq:curv_l2}
\end{equation}

\begin{equation}
||\kappa_{\Gamma}-\widetilde{\kappa}_{\Gamma}||_{\infty}=\max_{\Gamma}(|\kappa_{\Gamma_{k}}-\widetilde{\kappa}_{\Gamma_{k}}|),\label{eq:curv_linf}
\end{equation}

\begin{table}[H]
\begin{centering}
\begin{tabular}{|c|c|c|c|c|}
\hline 
 & \multicolumn{2}{c|}{${||\mathbf{u}-\mathbf{\tilde{u}}||}$ } & \multicolumn{2}{c|}{${||\mathbf{\kappa_{\textrm{\ensuremath{\Gamma}}}}-\widetilde{\kappa}_{\Gamma}||}$ }\tabularnewline
\hline 
\hline 
${\Delta}$  & ${L_{2}}$  & ${L_{\infty}}$  & ${L_{2}}$  & ${L_{\infty}}$ \tabularnewline
\hline 
\hline 
1/20  & 4.60E-9  & 1.50E-8 & 7.23E-8  & 1.27E-7\tabularnewline
\hline 
1/40 & 4.20E-9 & 1.11E-8 & 5.42E-8  & 1.26E-7\tabularnewline
\hline 
1/60 & 3.45E-9 & 8.76E-9  & 5.39E-8  & 9.32E-8 \tabularnewline
\hline 
1/80  & 3.34E-9 & 8.75E-9  & 4.59E-8 & 8.53E-8 \tabularnewline
\hline 
\end{tabular}
\par\end{centering}

\caption{${L_{2}}$ and ${L_{\infty}}$ error of velocity and curvature in
moving wedge for angle ${\theta=60}$ at ${t^{*}=1}$}\label{t3}
\end{table}

\noindent where the curvature is computed at the interface, which is interpolated
from level-set values. It can be seen that the error is very small
and decreases with grid refinement.

\subsection{Static Parabolic Interface}\label{2.4.4}

\begin{figure}
\centering{}\includegraphics[width=0.75\textwidth]{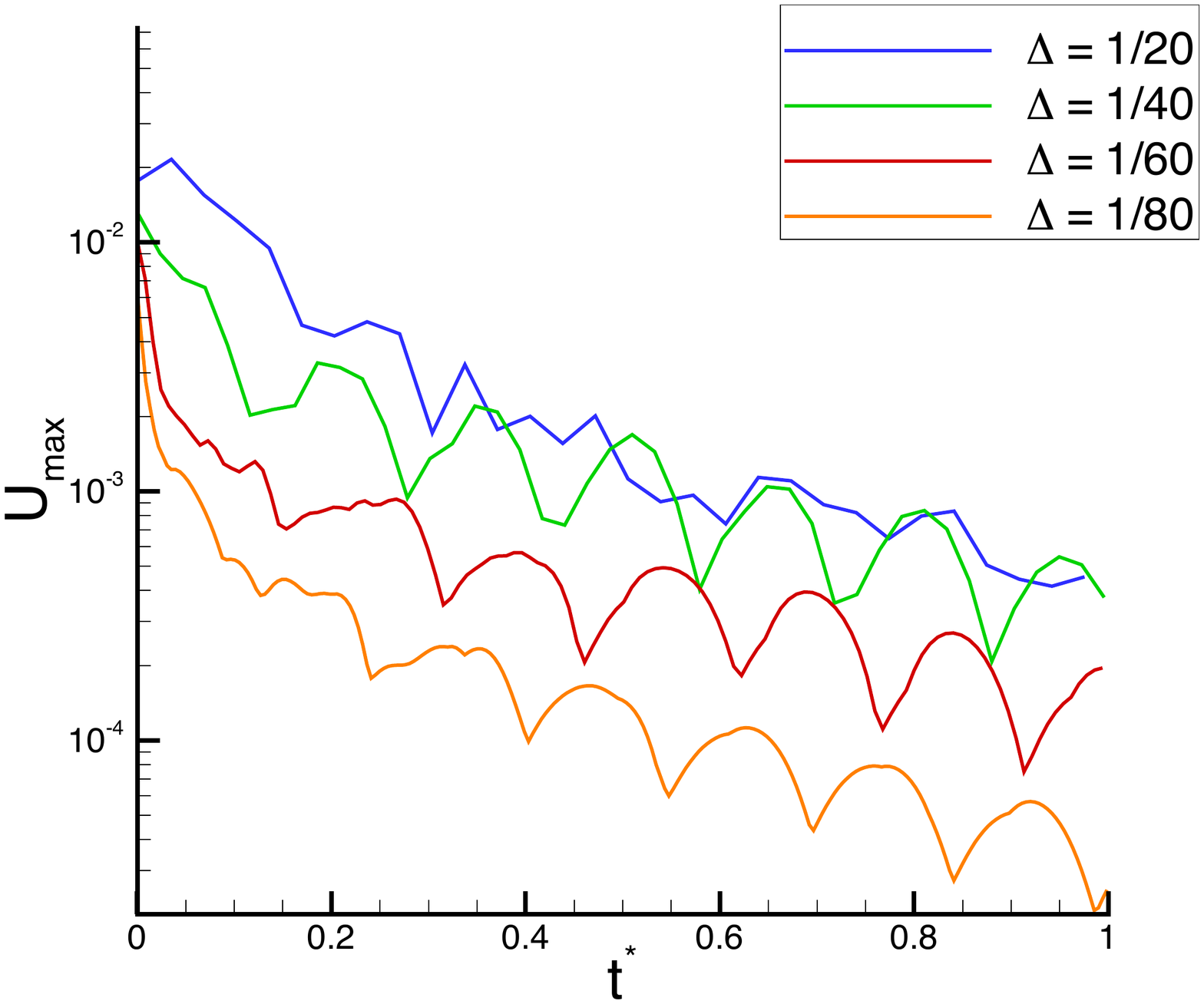}
\caption{Convergence of spurious currents for the static parabolic interface
intersecting the domain boundary at angle ${\theta=60}$}\label{fig:umx_staticint}
\end{figure}

\begin{figure}
\centering{}\includegraphics[width=0.65\textwidth]{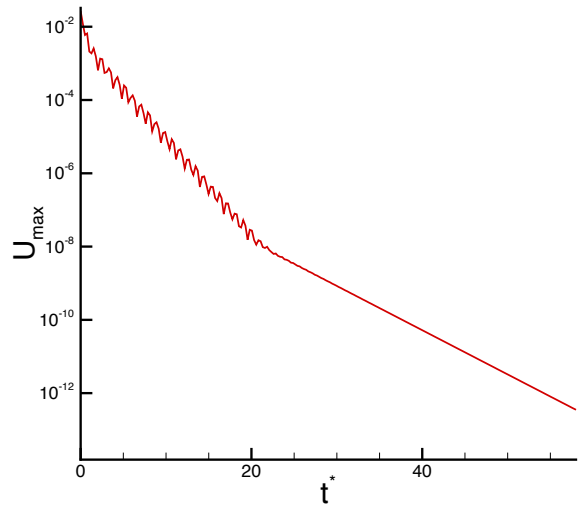}
\caption{Convergence of spurious currents to machine precision for the static
parabolic interface intersecting the domain boundary at angle ${\theta=60}$
(${\Delta=1/60}$)\label{fig:umx_staticint2} }
\end{figure}

The second verification test is the static fluid parabolic interface
intersecting the domain border at an imposed angle (see Figure \ref{fig:moving_int_streamlines}
for interface shape). There is no-flow at the left and right boundaries,
and free-slip is imposed at the bottom and top boundaries. Again,
the pressure and velocity fields are computed and the accuracy of
the method is characterized by the velocity and curvature error norms.
The Laplace number is chosen to characterize the flow behavior, where
${L}$ is a characteristic length scale equal to the interface radius
when ${\theta=60}$. In this test, La = 12000 which is similar to
the static droplet test case in section \ref{2.4.1}. Figure \ref{fig:umx_staticint} shows the evolution
of maximum velocity (scaled by ${U}$) with respect to the dimensionless
time $t^{*}$ for a circular interface at ${\theta=60}$. The capillary
waves converge with grid refinement, which is similar to the case
for the static droplet as illustrated above in Figure \ref{fig:umx_static_drop}. In addition,
The maximum velocity converges to machine precision as shown in Figure
\ref{fig:umx_staticint2}.  Tables \ref{t4} and \ref{t5} show the velocity, as well as the curvature error
norms for different imposed contact angles. The velocity converges
when the grid is refined for the considered contact angles. However,
the velocity error increases as the contact angle decreases because
the interface is more curved when the angle approaches zero and the
linear extension of the interface near the border becomes less accurate
\cite{spelt2005level}. Table \ref{t5} indicates that the curvature does not
converge with grid refinement because the linear extension of the
interface used for piecewise projection introduces a jump in curvature
at the interface. However, the error magnitude remains relatively
small.

\begin{table}[H]
\begin{centering}
\begin{tabular}{|c|c|c|c|c|c|c|}
\hline 
 & \multicolumn{2}{c|}{$\theta=60$} & \multicolumn{2}{c|}{$\theta=45$} & \multicolumn{2}{c|}{$\theta=30$}\tabularnewline
\hline 
\hline 
$\Delta$ & $L_{2}$ & $L_{\infty}$ & $L_{2}$ & $L_{\infty}$ & $L_{2}$ & $L_{\infty}$\tabularnewline
\hline 
\hline 
1/20 & 4.08E-4 & 1.28E-3 & 7.15E-4 & 2.16E-3 & 2.89E-3 & 1.50E-2\tabularnewline
\hline 
1/40 & 1.14E-4 & 3.51E-4 & 3.47E-4 & 1.20E-3 & 9.29E-4 & 3.95E-3\tabularnewline
\hline 
1/60 & 6.04E-5 & 2.14E-4 & 1.99E-4 & 8.35E-4 & 6.03E-4 & 2.29E-3\tabularnewline
\hline 
1/80 & 2.41E-5 & 7.35E-5 & 1.28E-4 & 4.45E-4 & 5.36E-4 & 2.23E-3\tabularnewline
\hline 
\end{tabular}
\par\end{centering}

\caption{Velocity ${L_{2}}$ and ${L_{\infty}}$ error in static parabolic
interface for different angles at ${t^{*}=1}$}\label{t4}
\end{table}
\begin{table}[H]
\begin{centering}
\begin{tabular}{|c|c|c|c|c|c|c|}
\hline 
 & \multicolumn{2}{c|}{$\theta=60$} & \multicolumn{2}{c|}{$\theta=45$} & \multicolumn{2}{c|}{$\theta=30$}\tabularnewline
\hline 
\hline 
$\Delta$ & $L_{2}$ & $L_{\infty}$ & $L_{2}$ & $L_{\infty}$ & $L_{2}$ & $L_{\infty}$\tabularnewline
\hline 
\hline 
1/20 & 3.09E-3 & 3.62E-3 & 5.11E-3 & 3.70E-3 & 5.66E-3 & 1.55E-2\tabularnewline
\hline 
1/40 & 9.84E-4 & 1.41E-3 & 1.93E-3 & 2.33E-3 & 4.37E-3 & 6.74E-3\tabularnewline
\hline 
1/60 & 8.02E-4 & 8.67E-4 & 1.22E-3 & 1.66E-3 & 5.56E-3 & 6.64E-3\tabularnewline
\hline 
1/80 & 4.78E-4 & 6.45E-4 & 1.80E-4 & 2.96E-4 & 4.98E-3 & 5.25E-3\tabularnewline
\hline 
\end{tabular}
\par\end{centering}

\caption{Curvature ${L_{2}}$ and ${L_{\infty}}$ error in static parabolic
interface for different angles at ${t^{*}=1}$}\label{t5}
\end{table}

\subsection{Moving Parabolic Interface}\label{2.4.5}

\begin{figure}[H]
\centering{}\includegraphics[width=0.65\textwidth]{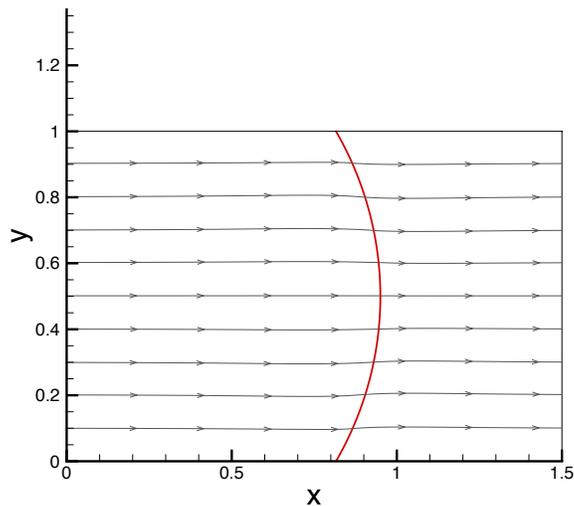}
\caption{Streamlines for the moving parabolic interface intersecting the domain
boundary at angle ${\theta=60}$\label{fig:moving_int_streamlines} }
\end{figure}

In this test, a parabolic interface is advected in a uniform constant
flow. There have been several attempts to analyze the accuracy of
the algorithm in terms of spurious currents for the translating droplet
\cite{Popinet2009,Abadie2015,Raeini2012}. However, to the best of
our knowledge, the translating parabolic interface with an imposed
contact angle coupled with Navier-Stokes has not been tested. The
static parabolic interface has been analyzed with the conventional
PDE-based reinitialization modified to consider contact angle \cite{DellaRocca2014},
but the magnitude of spurious currents was not reported. Figure \ref{fig:moving_int_streamlines}
shows that the flow streamlines are slightly bent around the interface
due to the presence of spurious currents. Table \ref{t6} shows the velocity
and curvature errors at ${t^{*}=1}$ with $La$ = 12000 and $We$
= 0.4. The numerical method converges upon grid refinement for velocity
as shown in Table \ref{t6}; however, the curvature error does not converge
with grid refinement. This is also due to the fact that the interface
is linearly extended below the computational domain. The maximum curvature
error has been computed without piecewise interface reconstruction,
which is shown in the last column of table \ref{t6}, and it converges with
grid refinement.

\begin{table}[H]
\begin{centering}
\begin{tabular}{|c|c|c|c|c|c|}
\hline 
 & \multicolumn{2}{c|}{${||\mathbf{u}-\mathbf{\tilde{u}}||}$  } & \multicolumn{3}{c|}{${||\mathbf{\kappa_{\textrm{\ensuremath{\Gamma}}}}-\widetilde{\kappa}_{\Gamma}||}$ }\tabularnewline
\hline 
\hline 
${\Delta}$  & ${L_{2}}$  & ${L_{\infty}}$  & ${L_{2}}$  & ${L_{\infty}}$  & ${L_{\infty}}$ without proposed reinitialization\tabularnewline
\hline 
\hline 
1/20  & 2.05E-3  & 1.14E-2 & 8.97E-3  & 1.80E-2 & N/A\tabularnewline
\hline 
1/40 & 9.62E-4  & 5.13E-3 & 5.93E-3  & 1.44E-2 & 1.08E-2\tabularnewline
\hline 
1/60 & 7.14E-4  & 2.83E-3  & 4.56E-3  & 1.66E-2  & 4.41E-3\tabularnewline
\hline 
1/80  & 6.11E-4 & 2.33E-3  & 5.11E-3 & 1.68E-2  & 2.46E-3\tabularnewline
\hline 
\end{tabular}
\par\end{centering}

\caption{${L_{2}}$ and ${L_{\infty}}$ error of velocity and curvature in
moving parabolic interface for angle ${\theta=60}$ at ${t^{*}=1}$}

\label{t6}
\end{table}

\subsection{Advancing interface in a capillary tube}\label{2.4.6}

% \ignore{(see figure 8)}

\begin{figure}
\centering{}%
\begin{tabular}{cc}
\includegraphics[width=0.38\paperwidth]{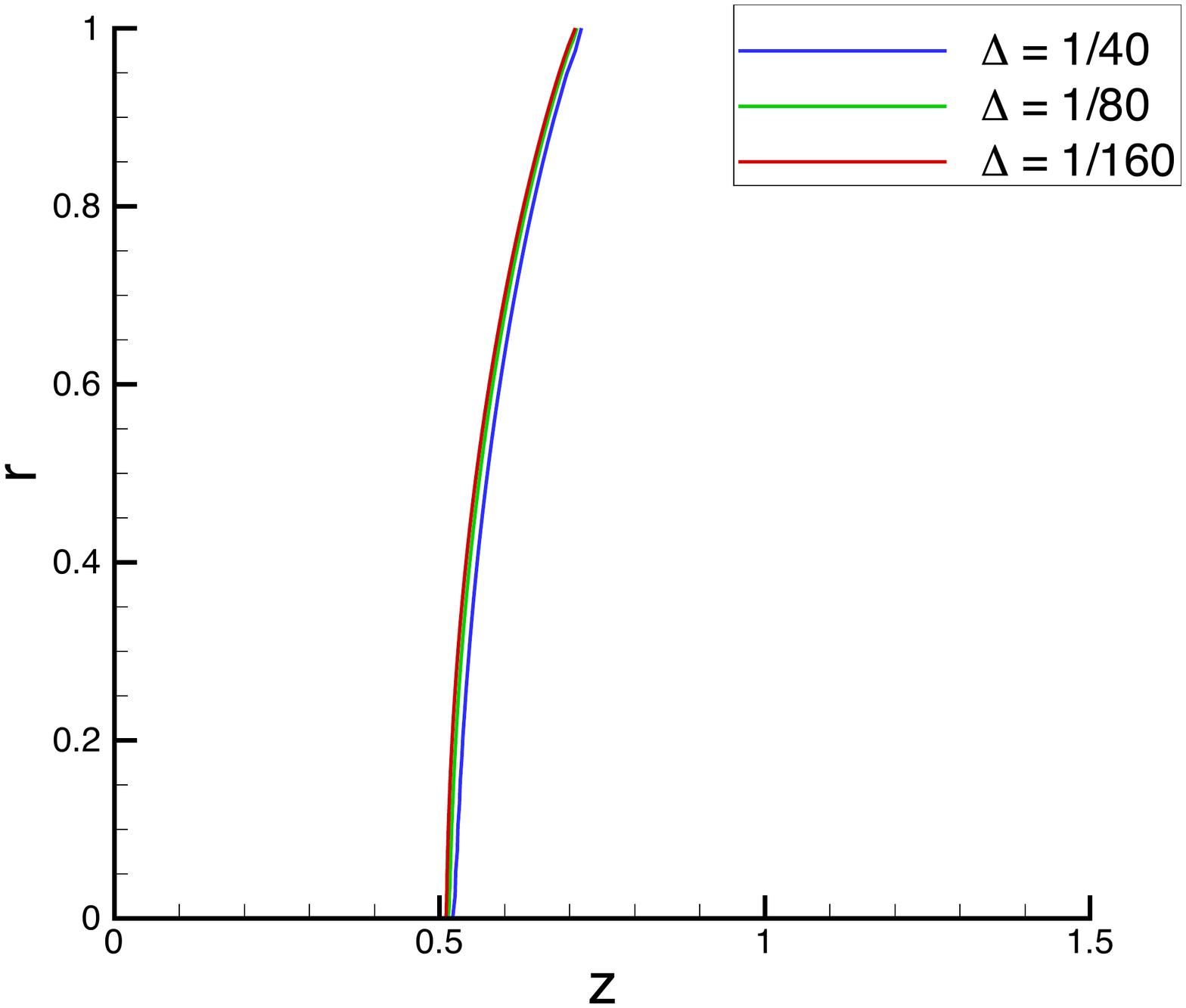} & \includegraphics[width=0.5\textwidth]{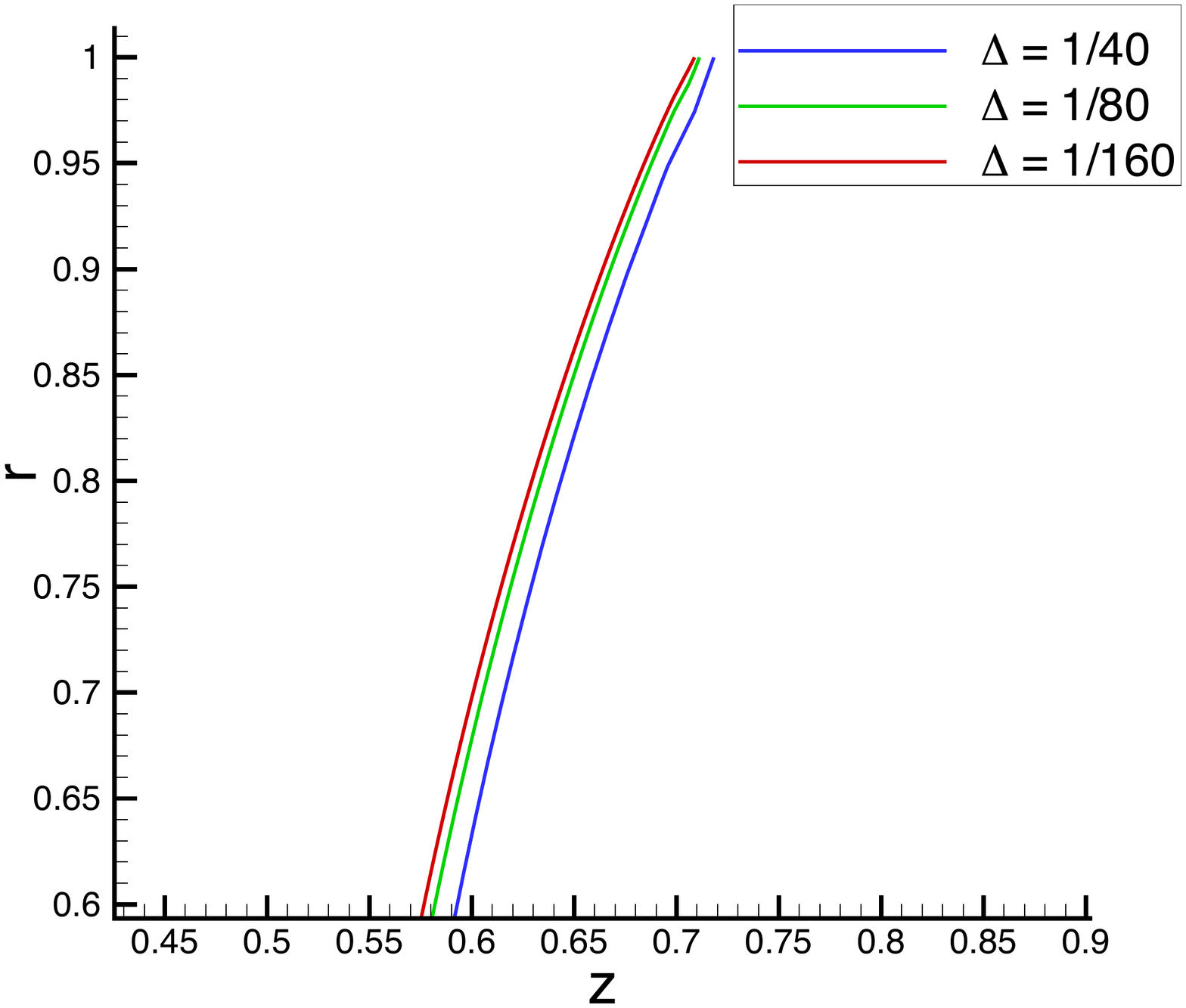}\tabularnewline
(a) The upper part of the steady-state interface & (b) Zoomed in near the contact angle\tabularnewline 
\end{tabular} \caption{Interface steady-state shape convergence for different grid sizes
for case 1}\label{fig:angle0}
\end{figure}

\begin{figure}
\centering \includegraphics[width=0.65\textwidth]{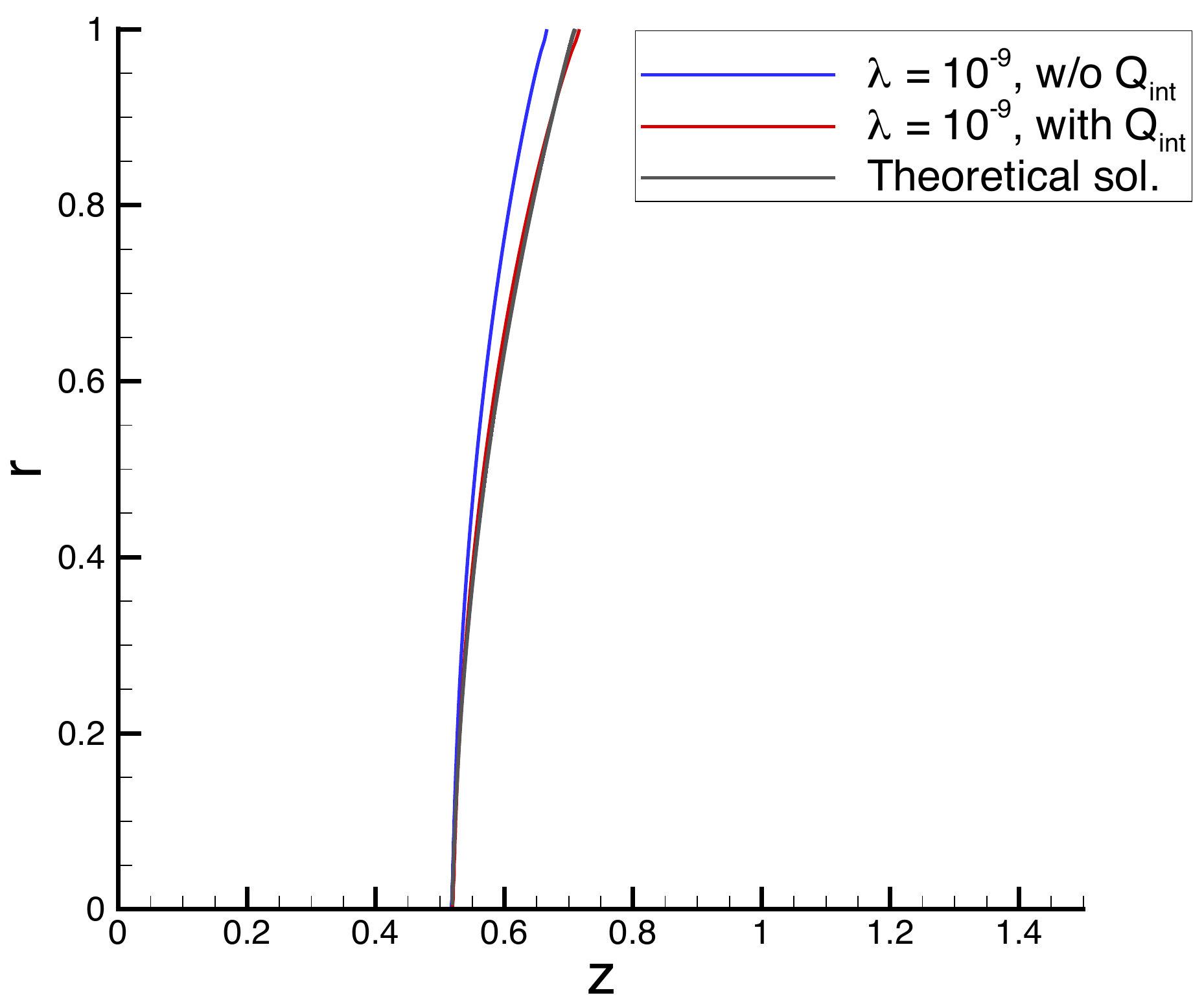}
\caption{The effect of the high order term ${Q_{int}}$ and comparison with
the theoretical solution for case 1}
\label{fig:higher_order0} 
\end{figure}

\begin{figure}
\centering \includegraphics[width=0.65\textwidth]{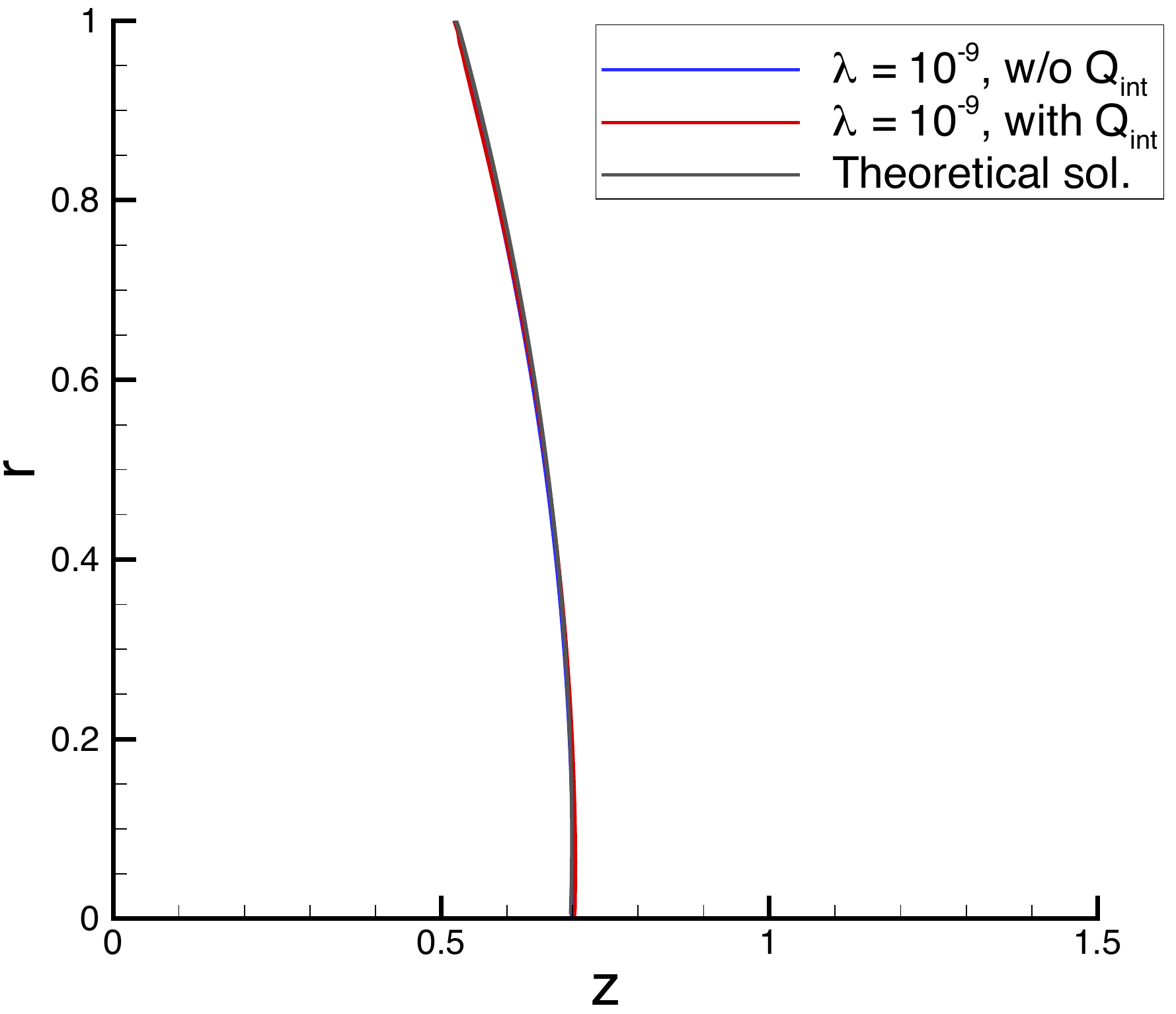}
\caption{The effect of the high order term ${Q_{int}}$ and comparison with
the theoretical solution for case 2}
\label{fig:higher_order60} 
\end{figure}

In this validation example, the displacement of a wetting fluid by
a non-wetting fluid in a capillary tube is investigated and compared
with experiments from \cite{Hoffman1975}. Also, the result is compared
with the solution of ordinary differential equation boundary-value-problem
derived for low capillary number steady-state interface in a capillary
tube \cite{RAME1996}. Table \ref{t7} lists parameters
for the two cases considered in \cite{Hoffman1975}. Both correspond
to different equilibrium contact angles. The radius of the capillary
tube is ${r=1}$ mm. Only half of the capillary tube is simulated,
and symmetry boundary condition is applied at the center of the tube.
The inlet boundary condition is a parabolic velocity (Poiseuille flow).
The computational domain size is ${1.5r}\times$ ${r}$, which is
found to be sufficient for a steady-state solution. The Cox-Voinov
dynamic contact angle model is implemented with Navier slip-length
of $\lambda={10^{-9}}$ m. The results are analyzed by considering
the convergence of the interface shape to the theoretical solution
of equation (1) in \cite{RAME1996}, which is:

\begin{equation}
\frac{{1}}{1-x}\frac{d[(1-x)\cos(\theta)]}{d(1-x)}=Ca\frac{{2\sin^{2}(\theta)}}{\theta-\sin(\theta)\cos(\theta)}\frac{1}{x}+B, \label{eq:diff_angle}
\end{equation}
where $\theta=\frac{\pi}{2}$ at $x=1$, and $\theta=\theta_{w}$
at $x=\lambda$. The parameter $B$ is a constant that is part of
the solution, and $x$ is the vertical distance from the wall to the
tube-center. The macroscopic contact angle, which is computed at a
vertical distance of 0.05 mm from the interface to the wall, is compared
as well. Figure \ref{fig:angle0} shows that the interface converges to a steady
state shape as the grid is refined for case 1. Figure \ref{fig:higher_order0} illustrates
the effect of the high order asymptotic term in Cox-Voinov asymptotic
theory. It can been seen that the high order term is required for
case 1 to match the theoretical solution. As for case 2, the high
order term corresponding to ${\theta_{w}=69}$ in Table 1 of Hocking \cite{Hocking1982}
is small (${Q_{int}=0.35}$) compared to the first case (${Q_{int}=-2.4}$).
Figure \ref{fig:higher_order60} shows a comparison between the numerical solution (with
and without the higher order correction term) with the theoretical
solution. The difference between the shapes is almost negligible.

% From equation 9, it can be seen that the Cox-Voinov model applied here is mesh dependent leading to grid-dependent numerical dynamic contact angle. Therefore, the results are also analyzed through the convergence of the average capillary pressure, which is related to a specific macroscopic contact angle via:%\begin{equation}%P_c = \frac{2 \sigma \cos \theta_m}{R}%\end{equation}

The computed macroscopic contact angle is compared with the observed
experimental results in \cite{Hoffman1975}. The last two columns
in Table \ref{t7} include the computed macroscopic angle and the observed
experimental macroscopic contact angle. The difference between the
computed and experimental macroscopic is around the 5\% error measurement
mentioned in \cite{Hoffman1975}.% \ignore{The numerical intermediate
%contact angle approaches the constant microscopic contact angle logarithmically,
%which is expected since the The computed capillary pressure and macroscopic
%contact angle converge with grid refinement, and it is found to be
%within the 5 error measurement mentioned in \cite{Hoffman1975}.}
%\ignore{In addition, the difference between the numerical intermediate
%contact angle () converges logarithmically to the microscopic contact
%angle cubed () (equal to the static angle in the model), which is
%consistent with the Cox-Voinov theory.}

\begin{table}
\begin{centering}
\begin{tabular}{|c|c|c|c|c|}
\hline 
 & Ca & $\theta_{eq}$ & $\theta_{macro}$ sim. & $\theta_{macro}$ exp.\tabularnewline
\hline 
\hline 
Case 1 & 1.83E-2 & 0 & 67.6 & 66.5\tabularnewline
\hline 
Case 2 & 4.94E-2 & 69 & 107.1 & 114\tabularnewline
\hline 
\end{tabular}
\par\end{centering}

\caption{Simulation and experimental macroscopic contact angle for two different
cases in \cite{Hoffman1975}}

\label{t7}
\end{table}

\subsection{Capillary rise}\label{2.4.7}

\begin{figure}[H]
\centering{}%
\begin{tabular}{ccc}
 \includegraphics[scale=0.25]{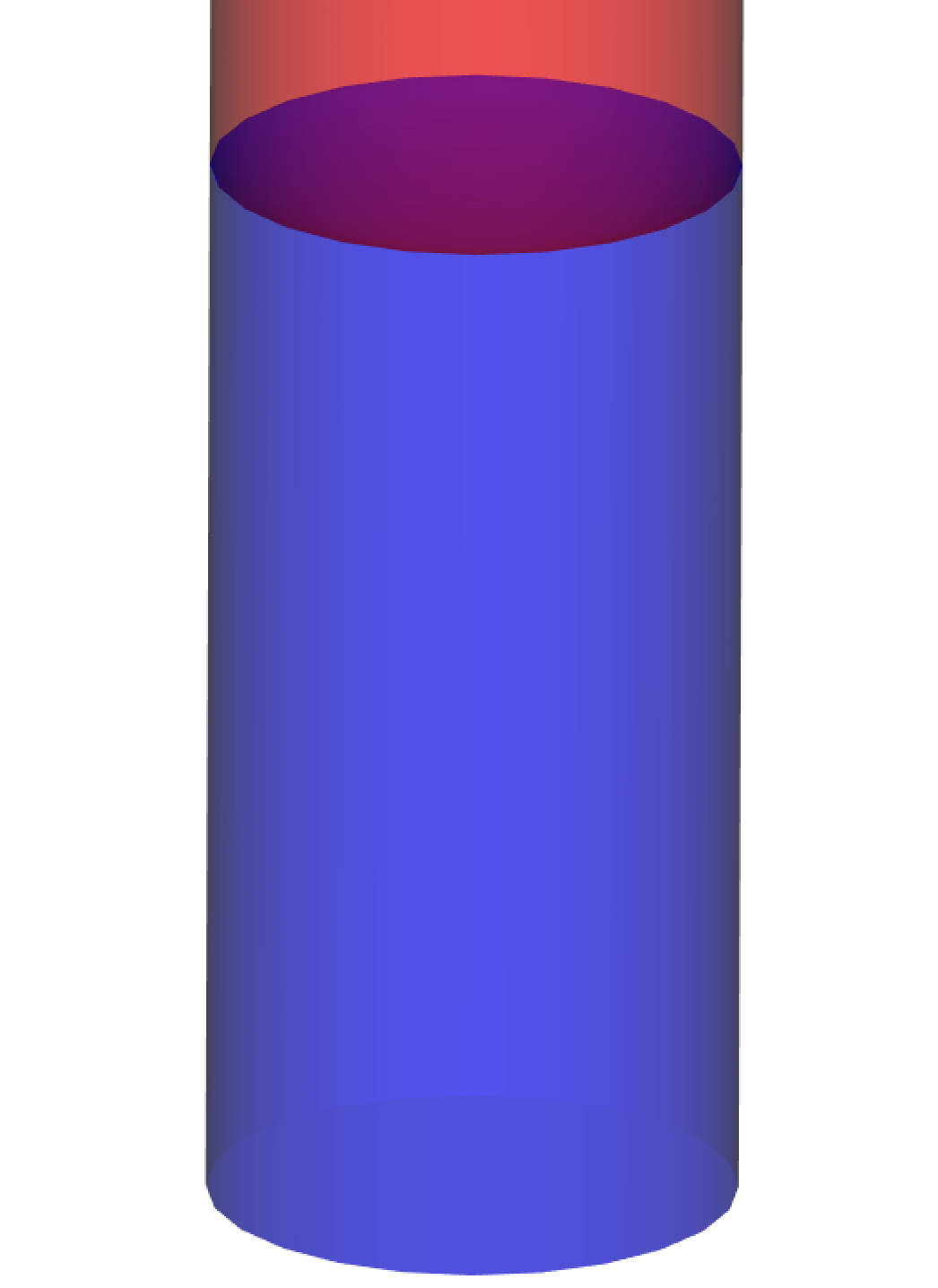} &  \includegraphics[scale=0.25]{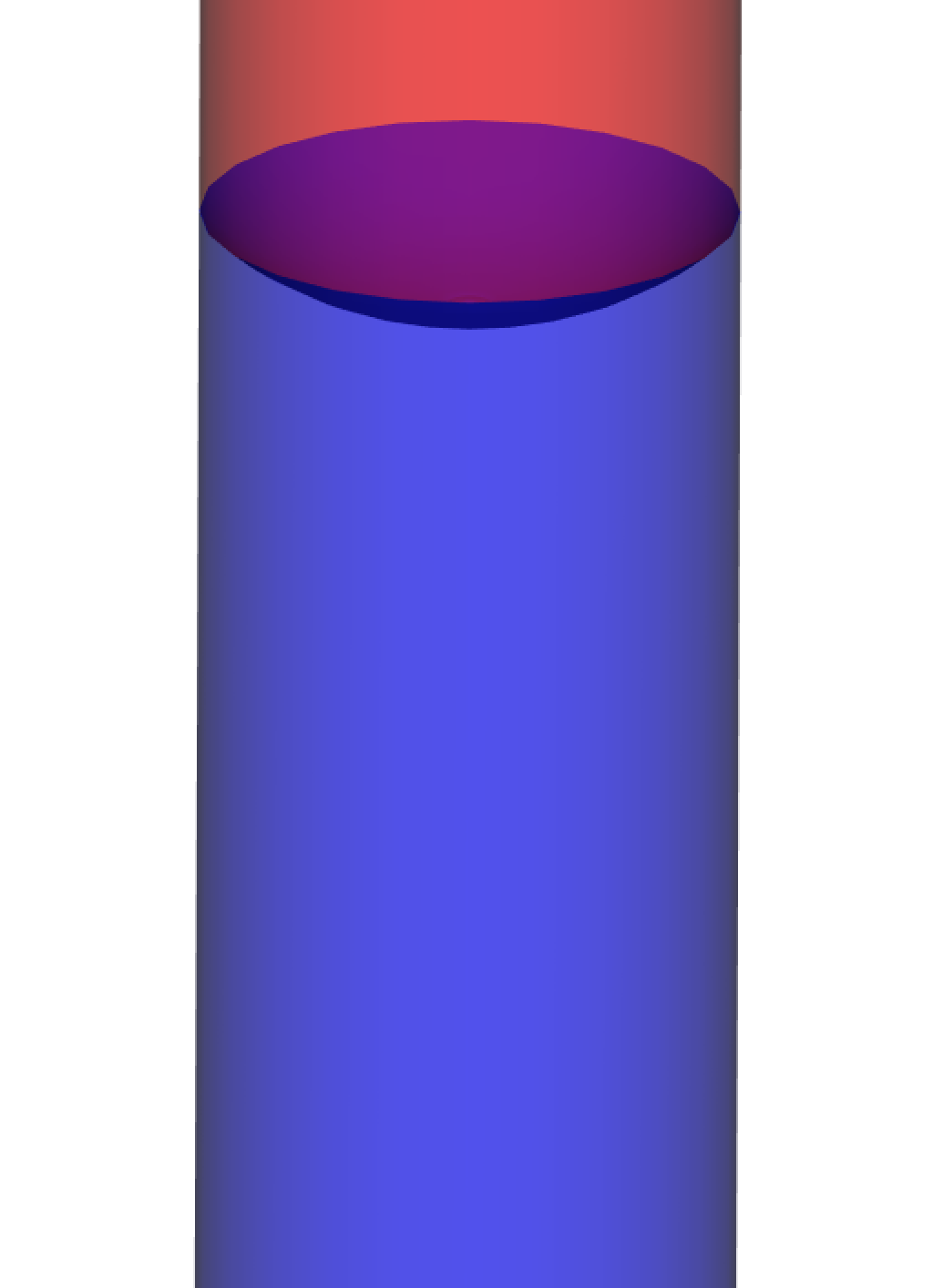} & \includegraphics[scale=0.25]{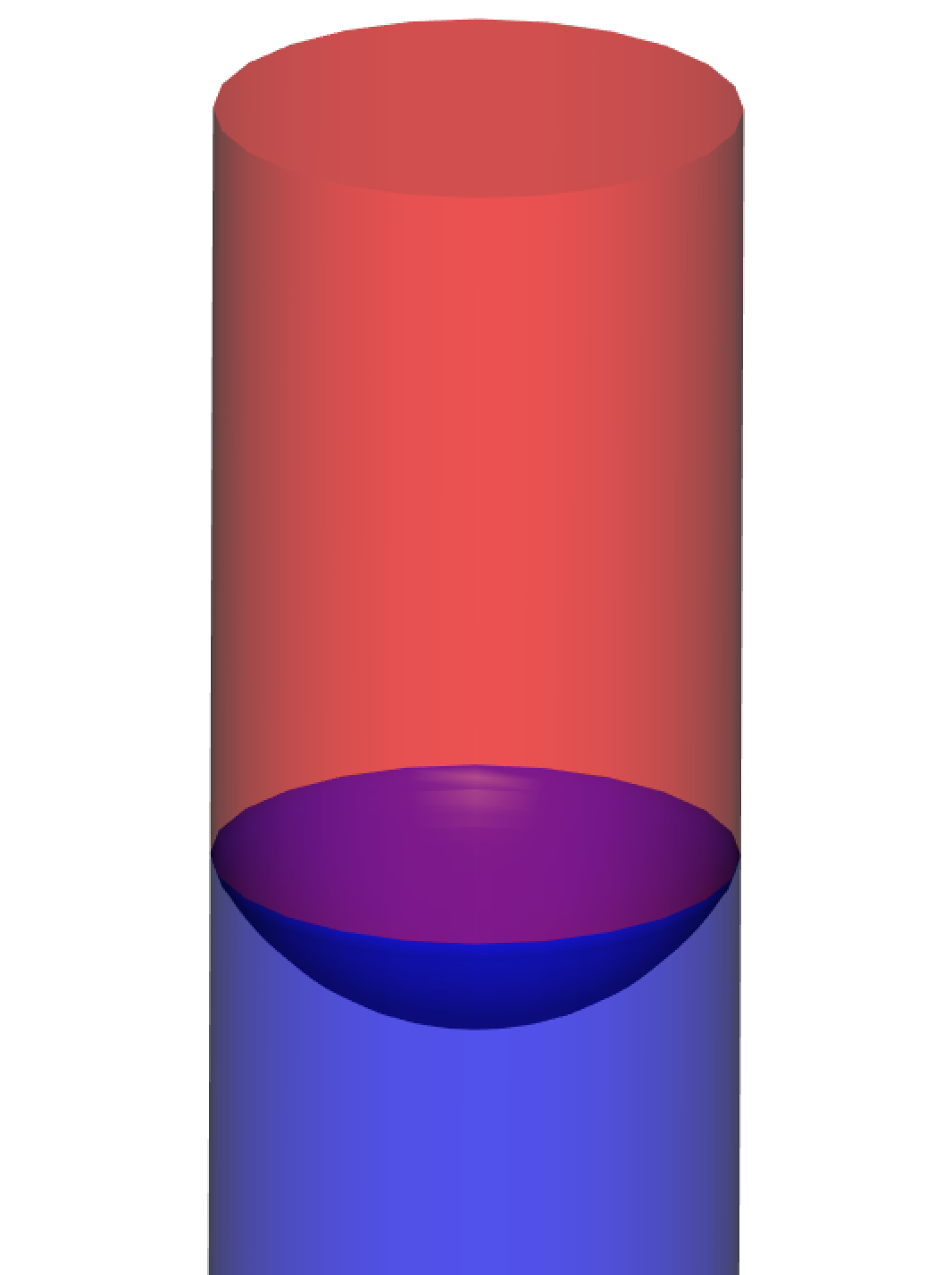}\tabularnewline
(a) The most deformed  & (b) The deformed interface  & (c) The final equilibrium interface \tabularnewline
interface at t = 0.01 second  & at t = 0.05 seconds & shape at t = 0.4 seconds\tabularnewline
 &  & \tabularnewline
\end{tabular} \caption{Interface shape at different stages (${\Delta=1/20}$)}
\label{fig:Interface_h}
\end{figure}

\begin{figure}[H]
\centering{} \includegraphics[width=0.75\textwidth]{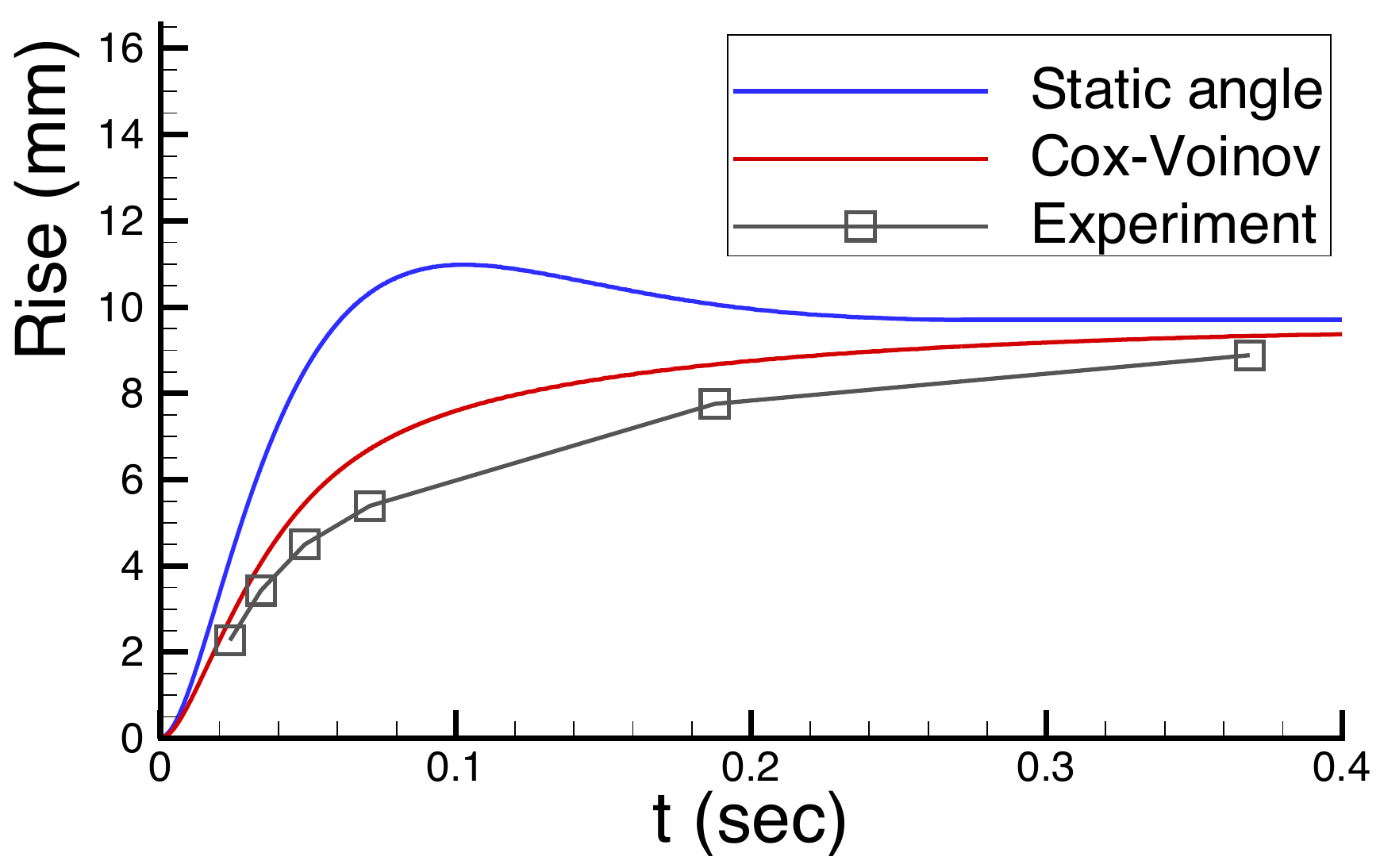}
\caption{Average interface height vs. time for both static and dynamic angle
(${\Delta=1/20}$) and comparison with experiment \cite{Heshmati2013}} \label{fig:capillary_rise}
\end{figure}

In this example, our new method is used to model meniscus rise in
a vertical tube. When a vertical capillary tube is brought in contact
with a wetting fluid, the fluid spontaneously rises through the tube.
The final location of the fluid interface is reached whenever the
capillary pressure is balanced by gravity:

\begin{equation}
P_{c}=\frac{2\gamma\cos\theta_{eq}}{R}=\rho gh,\label{eq:rise}
\end{equation}
where ${h}$ is the final height of the fluid interface. The dynamics
of the fluid interface between the initial fluid tube contact and
the capillary-gravity balance can only be predicted accurately by
modeling the correct physics of contact line dynamics. The simulation
results are compared with recent experiment in \cite{Heshmati2013},
and Table \ref{t8} shows the fluid parameters for the relevant case. Symmetry
boundary condition is imposed at the center of the tube, and atmospheric
pressure boundary conditions are applied at the inlet and outlet of
the domain. The Cox-Voinov model with slip-length ${\lambda=10^{-9}}$
m is used for the moving contact line. Figure \ref{fig:Interface_h} illustrates the interface
shape when it is rising at two different times as well as when it
reaches the final equilibrium state. The time evolution of the average
fluid interface is compared with data taken from Figure 5 in \cite{Heshmati2013}.
In addition, the interface height is also tracked when a constant
contact angle boundary condition is applied instead of the Cox-Voinov
model. From Figure \ref{fig:capillary_rise}, it is seen that the numerical method with Cox-Voinov
model predicts the trend of experimental measurements with slight
difference. The discrepancy occurs at the beginning, and most likely
this is when inertia plays a role in the dynamics where the Cox-Voinov
model does not take into account. On the other hand, applying a constant
contact angle without resolving the slip length results in an interface
overshoot above the equilibrium height, which is inaccurate and unphysical.

\begin{table}[H]
\begin{centering}
\begin{tabular}{|c|c|c|c|c|}
\hline 
$\theta_{eq}$ & R (mm) & $\rho$ (kg/m$^3$) & $\mu$ (cP) & $\gamma$(mN/m)\tabularnewline
\hline 
\hline 
0 & 0.65 & 744 & 2.6 & 24.8\tabularnewline
\hline 
\end{tabular}
\par\end{centering}

\caption{Fluid properties corresponding to Soltrol 170 \cite{Heshmati2013}}\label{t8}
\end{table}

\subsection{Forced imbibition of viscous oil by water}\label{2.4.8}

\begin{figure}
\centering{}%
\begin{tabular}{cc}
\includegraphics[width=0.3\paperwidth]{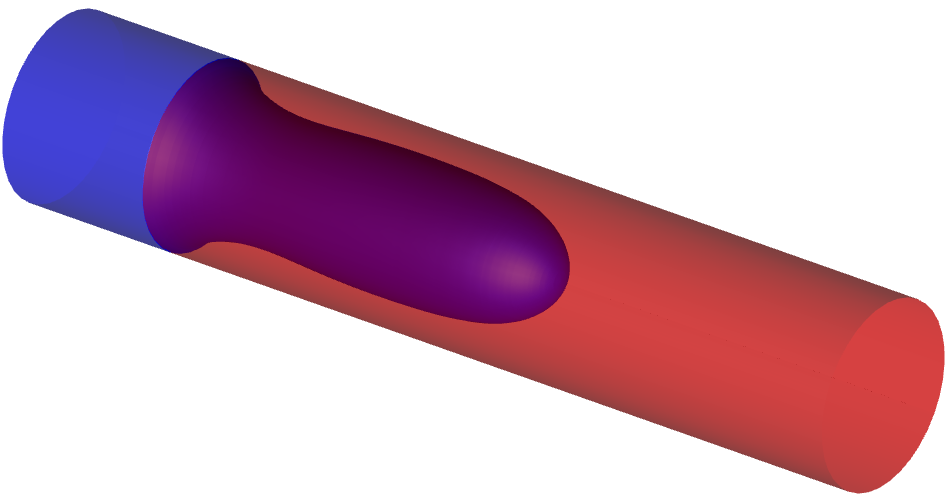} & \includegraphics[width=0.45\textwidth]{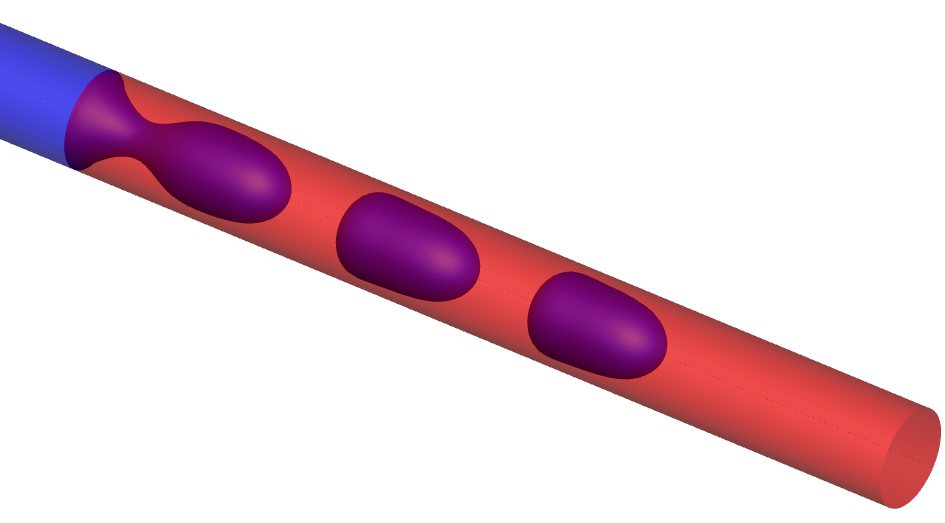}\tabularnewline
(a) Case 1: Ca = 1.14E-3 & (b) Case 2: Ca = 4.02E-4\tabularnewline 
\end{tabular} \caption{The water/oil interface shape that is deformed due to viscous forces at different capillary numbers}\label{fig:Toong_3D}
\end{figure}

In this validation example, the forced displacement of a very viscous fluid inside a capillary tube is simulated. The results are compared with experiments by Hansen and Toong \cite{hansen1971interface}, where they showed schematics of the water/viscous oil interface shape at different capillary numbers. The wetting water displaces the non-wetting viscous oil, and large viscous bending occurs in the intermediate region. In addition, the displacement process is likely to become unstable because low viscous fluid (water) is injected into a higher viscous fluid (paraffin oil). Table \ref{t8_2} shows the parameters used to simulate this forced imbibition process. The contact angle has not been reported in \cite{hansen1971interface}. They only mention that the capillary tube is more wetting towards the water. In the simulation, the contact angle is assumed to be $\theta_{eq}=70^o$ with respect to the water, which makes it the wetting fluid. Figure \ref{fig:Toong_3D} shows the water/viscous-oil interface at two different capillary numbers. For case 1, the capillary number is Ca = 1.14E-3, where the viscosity of the injected fluid is used. The capillary number is sufficient enough to bend the fluid interface, and viscous fingering occurs as shown in Figure \ref{fig:Toong_3D} (a). The results agree with the interface shape reported in the experiment at Ca = 1.14E-3 (Figure 6(d) in \cite{hansen1971interface}). The computed macroscopic contact angle based on the Cox-Voinov theory is $\theta_{macro} = 120^o$.

\begin{figure}
\centering{}%
\begin{tabular}{cc}

\tabularnewline

\tabularnewline
\includegraphics[width=0.33\paperwidth]{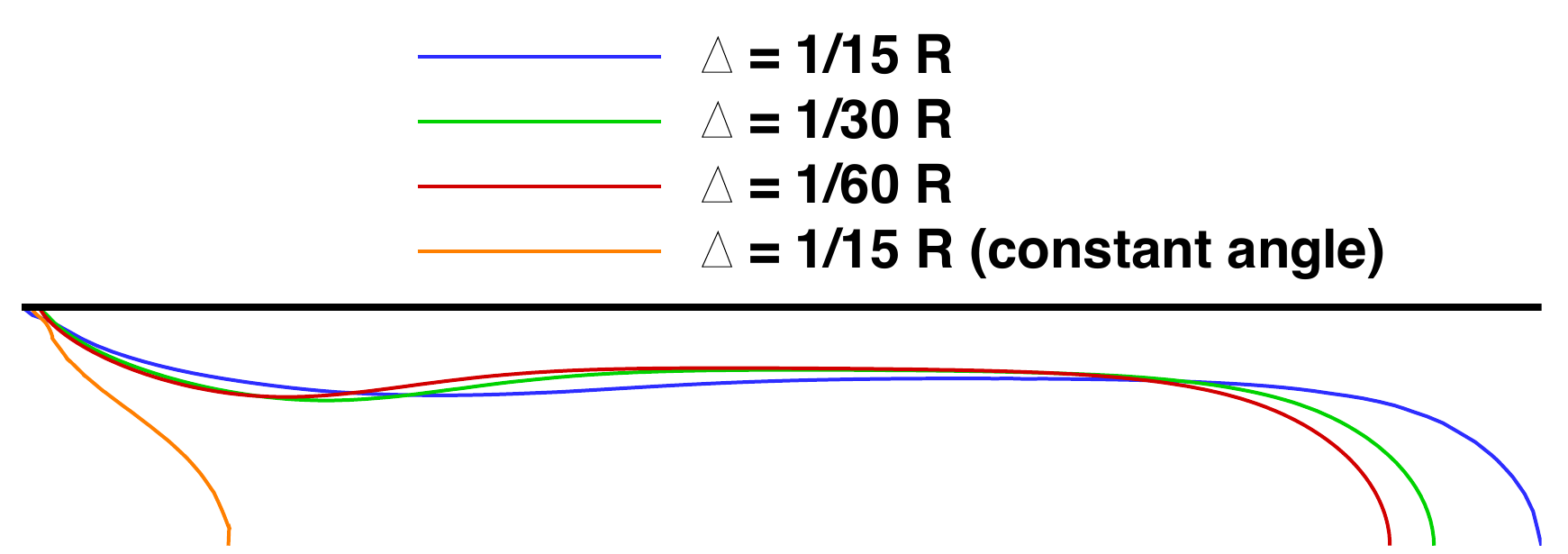} & \includegraphics[width=0.5\textwidth]{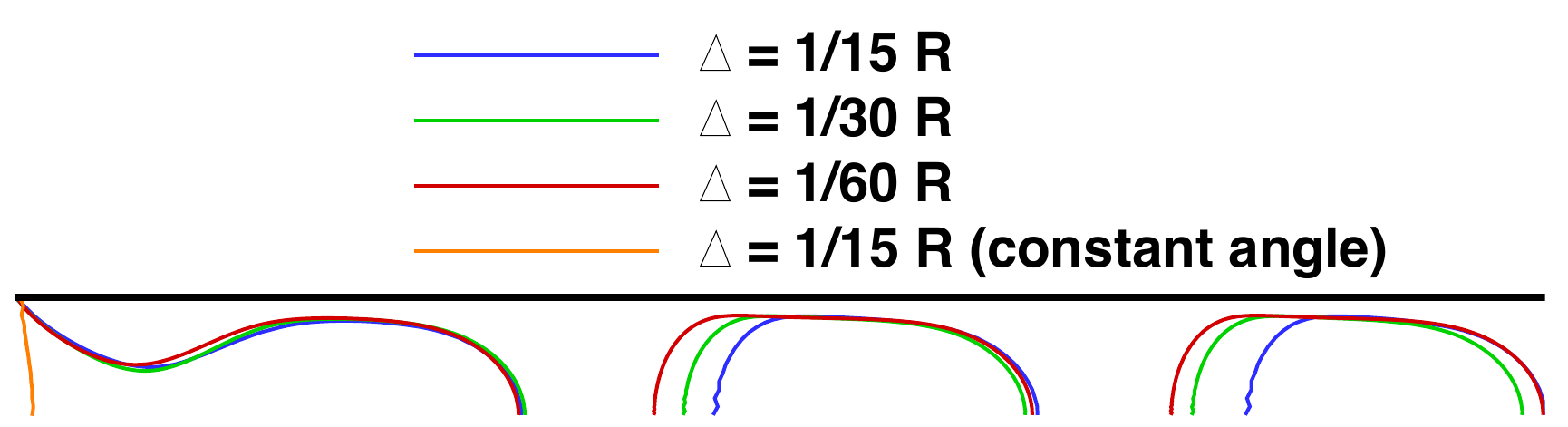}\tabularnewline
(a) Case 1: Ca = 1.14E-3 & (b) Case 2: Ca = 4.02E-4\tabularnewline 
\end{tabular} \caption{Interface shape convergence for different grid sizes}\label{fig:Toong_convergence}
\end{figure}

At Ca = 4.02E-4, the capillary force competes with the viscous force due to water injection. The viscous force bends the interface towards the oil side. However, the capillary force and the water wettability prefers to lower the system energy by minimizing the interfacial area. Therefore, the water/oil interface breaks into water droplets. The interface breakup is periodic as seen in Figure \ref{fig:Toong_3D} (b). This complex interfacial dynamics of generating periodic droplets has also been reported in \cite{hansen1971interface} at the same capillary number of Ca = 4.02E-4. The macroscopic contact angle in this case is $\theta_{macro} = 115^o$.  Figure \ref{fig:Toong_convergence} shows the convergence of the simulation results. It is also confirmed in this example that the Cox-Voinov model is necessary to match the experimental results. Applying a constant contact angle, without resolving the slip-length, results in a completely different fluid interface shape for both capillary numbers as shown in the black curves in Figure \ref{fig:Toong_convergence} .

\vspace{0mm}

\begin{table}[H]
\begin{centering}
\begin{tabular}{|c|c|c|c|c|}
\hline 
$\theta_{eq}$ & R (mm) & $\rho_w/\rho_o$ (kg/m$^\text{3}$) & $\mu_w/\mu_o$ (cP) & $\gamma$(mN/m)\tabularnewline
\hline 
\hline 
70 & 1.23 & 993/874 & 0.99/177 & 51.9\tabularnewline
\hline 
\end{tabular}
\par\end{centering}

\caption{Fluid properties corresponding to water and paraffin oil (Nujol). The parameters are taken from \cite{hansen1971interface}}\label{t8_2}.
\end{table}
\section{Conclusion}\label{2.5}

A new numerical method based on linear reconstruction of the interface
has been developed to simulate two-phase flow with moving contact
lines. The method is accurate in terms of representing the fluid interface. In terms of spurious currents, the velocity error converges
to machine precision for static interfaces that include contact angle
boundary condition and the error decreases with grid refinement when
the interface is advected in a constant velocity field. The method
combined with the Cox-Voinov model is validated and compared with
the theory for advancing interface in a capillary tube. The higher-order
asymptotic term in method of matched asymptotic expansions in the
Cox-Voinov model is necessary to match the theoretical solution for
small contact angles. In addition, the method has been compared with
experiments for both advancing fluid in a capillary tube and vertical
capillary rise which shows good agreement. %as well as oscillating droplet, which show very go

\section*{}

\section*{Acknowledgments }

We thank Saudi Aramco and SUPRI-B at Stanford University for funding
this research. We also thank Shahriar Afkhami and St{\'e}phane Zaleski for useful
discussions.

%The contribution of the method is two folds. First, the method
%is computationally more efficient than the standard PDE-based reinitialization
%as well as more accurate in representing the energy of droplet oscillations
    
\bibliographystyle{acm}
%\nocite{*}
\bibliography{Bibliography_Full}

\end{document}